\documentclass[prd,notitlepage,nofootinbib,superscriptadress,11pt]{revtex4-1}
\pdfoutput=1
\usepackage[english]{babel}
\usepackage{amsmath,amssymb,amsfonts, bm,bbm,slashed, subdepth}

\usepackage{graphicx}
\usepackage[sort&compress]{natbib}
\usepackage{xcolor}
\usepackage[normalem]{ulem}
\usepackage{hyperref}
\usepackage{cleveref}
\definecolor{red}{rgb}{1.0, 0, 0}
\usepackage{hyperref}
\usepackage{enumerate}
\usepackage{epsfig, subfigure}
\usepackage{booktabs, tabularx}
\usepackage{units}
\usepackage{bbold}
\usepackage{caption}

\newcommand{\be}{\begin{equation}}
\newcommand{\ee}{\end{equation}}
\newcommand{\ba}{\begin{array}}
\newcommand{\ea}{\end{array}}
\newcommand{\bea}{\begin{eqnarray}}
\newcommand{\eea}{\end{eqnarray}}
\newcommand{\balg}{\begin{align}}
\newcommand{\ealg}{\end{align}}
\newcommand{\bit}{\begin{itemize}}
\newcommand{\eit}{\end{itemize}}
\newcommand{\trm}[1]{\textrm{#1}}

\newcommand{\sv}{\langle \sigma_{\rm ann} v \rangle}
\newcommand{\psdphi}{{\cal F}_{\! \phi}}
\newcommand{\Gphi}{{\cal G}_{\phi}}
\newcommand{\Dpos}{{\cal D}_{e}}
\newcommand{\Dgam}{{\cal D}_{\gamma}}

\newcommand{\Mpc}{\trm{\Mpc}}
\newcommand{\yr}{\trm{\yr}}
\newcommand{\eV}{\trm{\eV}}

\begin{document}
\title{Dark matter indirect  signals with long-lived mediators}

\author{Xiaoyong Chu,}
\email{xiaoyong.chu@oeaw.ac.at}
\author{Suchita Kulkarni,}
\email{suchita.kulkarni@oeaw.ac.at}
\affiliation{Institut f\"ur Hochenergiephysik,  
\"Osterreichische Akademie der Wissenschaften, \\ Nikolsdorfer Gasse 18, 1050 Wien, Austria}
\author{Pierre Salati}
\email{pierre.salati@lapth.cnrs.fr}
\affiliation{LAPTh, Universit\'e Savoie Mont Blanc \& CNRS,\\
9 Chemin de Bellevue, B.P.110 Annecy-le-Vieux, F-74941 Annecy Cedex, France}

\begin{flushright}
HEPHY-PUB 986/17\\
LAPTH-020/17
\end{flushright}

\begin{abstract}
\vskip1.0cm
Dark matter particles could annihilate into light and metastable mediators subsequently decaying far away from where they are produced. In this scenario, the indirect signatures of dark matter are altered with respect to the conventional situation where standard model particles are directly injected where annihilation happens.
We revisit the long-lived particle proposal (Rothstein et al., 2009) and devise the tools to explore this new phenomenology. We calculate the effective dark matter distribution resulting from the smearing by mediator propagation. We derive general expressions for the fluxes of mediators and their decay particles. We study how the $J$-factor, which naturally appears in the calculation of the dark matter induced gamma ray signal, is modified in the presence of mediators. We also derive the anisotropy which the cosmic ray positron flux exhibits in this scenario.
We finally comment upon a recent proposal based on long-lived mediators where the effective dark matter density at the Earth is increased such as to explain the cosmic ray positron anomaly. We conclude that this scenario is barely tenable as regards the very dense dark matter spike which it requires at the Galactic center. The associated positron anisotropy is very small and undetectable, except at high energies where it reaches a level of order $10^{-4}$ to $10^{-3}$.
\end{abstract}

\maketitle
\newpage
\section{Introduction}
\label{sec:introduction}
The quest for understanding the origin and particle nature of Dark Matter (DM) inevitably leads to  formulation of New Physics (NP) scenarios. Despite intense searches and absence of a concrete signal, the Weakly Interacting Massive Particle (WIMP) paradigm remains to be one of the most attractive DM scenarios~\cite{Lee:1977ua, Steigman:1984ac}. Several NP scenarios can realize WIMP DM within which the DM relic density is generated via annihilation into other new or Standard Model (SM) particles~\cite{Srednicki:1988ce,Gondolo:1990dk}. Due to particle physics properties, these new particles (mediators) can be long-lived. Depending on their lifetimes, decays of these long-lived mediators to SM particles can be probed at several different experiments. In this paper, we consider the prospects for indirect detection of long-lived mediators.

Long-lived particles are well known in particle physics. They can manifest in the form of particle physics candidates such as the dark photon, dark Higgs or supersymmetric particles~\cite{Essig:2013lka}. Several model independent DM scenarios also feature long-lived particles~\cite{Agashe:2014yua, Pappadopulo:2016pkp}. The boosted DM scenario, e.g.~\cite{Agashe:2014yua}, assumes DM annihilation into a stable, light and correspondingly relativistic particle and explores the direct detection prospects of such light relativistic particle. If this particle is metastable instead, it can be probed at indirect detection experiments as well. 

As DM annihilates in the Galaxy, the annihilation products, namely gamma rays, neutrinos or charged leptons, can be detected at the Earth as a continuum or spectral lines. They can also lead to anisotropies in case their arrival direction is not isotropic. Several dedicated experiments look for such an excess of continuum/spectral lines and anisotropies, such as Fermi-Lat~\cite{Ackermann:2010ip}, H.E.S.S.~\cite{Abdalla:2016olq} and AMS~\cite{Aguilar:2013qda} collaborations. This forms the basis for indirect detection searches, as recently summarized in \cite{Cirelli:2016ony, Gaskins:2016cha}. Theoretical predictions for signals at indirect detection experiments depend on the underlying particle physics model, the DM density at the annihilation point and the understanding of cosmic ray propagation.

In case DM particles annihilate into long-lived mediators, the situation changes. Mediators introduce two essential differences with respect to the conventional situation where SM species are directly produced by DM annihilation.
To commence, mediators may decay very far away from where they are produced. The injection rate of SM particles is no longer directly related to the DM density. This opens the possibility to have large DM concentrations, say at the Galactic center (GC), which are no longer associated to strong signals. As the effective DM density that enters in the indirect signals is smeared by mediators, observations which so far were in tension with the existence of DM spikes are much less constraining. In the pioneering analysis by~\cite{Rothstein:2009pm}, based on non-relativistic long-lived mediators, observations of the gamma ray emission from the GC no longer preclude the large values of the WIMP annihilation cross section required to explain the cosmic ray positron excess.
The other difference lies in the small mass and high energy of the mediators. In general, these particles are relativistic at the time of annihilation and lead to anisotropic distributions of cosmic rays. In this paper, we set up the formalism for computing the fluxes and associated anisotropies of the prompt species -- positrons and photons -- produced in the decays of these long-lived mediators.

The paper is organized as follows.
In section~\ref{sec:eff_DM_rho}, we show how the DM density profile is smeared by mediators and compute the production rate of SM particles yielded by mediator decay. The conventional situation is recovered if the DM density is replaced by an effective value which depends on the mediator properties. In the scenario proposed by~\cite{Kim:2017qaw} to explain the cosmic ray positron excess, the effective DM density at the Earth is exceedingly large with respect to its actual value. The price to pay for such an enhancement is the existence of a very dense DM spike at the GC. In section~\ref{sec:blackhole} we ponder upon possible astrophysical mechanisms which can alter the Galactic density profiles and show that adiabatic contraction resulting from the formation of the GC black hole can marginally lead to the desired DM density.
In section~\ref{sec:mediated_DM_ID}, we build the formalism for computing the fluxes of mediators and prompt decay particles.  The observable flux of gamma rays is then described by an effective $J$-factor, which,  for very long-lived mediators, considerably differs from the canonical one. We carry on a general analysis assuming two-body decays and pay particular attention to the mediator velocity.
Section~\ref{sec:anisotropy} is devoted to the anisotropy which high-energy positrons exhibit in the presence of mediators decaying inside the cosmic ray ``last scattering'', or ``ballistic'', sphere. The formalism to compute it is presented and subsequently applied to the cases of two-body and three-body decays. Depending on the DM central profile, the positron anisotropy may reach a level of $10^{-4}$ to $10^{-3}$ at the TeV scale.
We finally summarize our results and conclude in section~\ref{sec:conclusion}.

\section{Mediator smeared Dark Matter profile}
\label{sec:medprof}

Throughout this article, the DM species $\chi$ pair annihilate into light mediators $\phi$ which, unless otherwise stated, are assumed (i) to be weakly interacting, (ii) to be much lighter than their progenitors $\chi$ and (iii) very long-lived so as to decay at Galactic distances from where they are produced. The DM mass $m_{\chi}$ has a benchmark value of ${\cal{O}}(1)$ TeV while the mediator mass is ${\cal{O}}(100)$ MeV. Mediators subsequently decay into light SM species like $e^{+}e^{-}$ pairs or photons, hence the chain of reactions
\be
\chi \, + \, \chi \, \to \, \phi \, + \, \phi
\;\;\;{\rm and}\;\;\;
\phi \, \to \, {\rm SM} \;.
\label{eq:chain}
\ee
In the former case, we can safely assume the hierarchy $m_{e} \ll m_{\phi} \ll m_{\chi}$ although our formalism is quite general and can be readily applied to more complicated situations.

In this section, we would like to derive the rate $q_{\rm \, SM}$ at which SM particles are supplied at location $\vec{x}$ by the decays of mediators initially produced at point $\vec{x}_{S}$ through DM annihilation. The two-step process~(\ref{eq:chain}) requires that we model the production of mediators as well as their subsequent propagation throughout the Galaxy. For illustration purposes, we assume that each mediator decay yields either two photons or a single positron, which we consider here to be the SM species of interest.

\subsection{Effective Dark Matter density}
\label{sec:eff_DM_rho}

To commence, DM particles pair annihilate with a cross section whose average over the initial momenta is denoted by $\sv$. In the case of Majorana DM, the annihilation rate includes a statistical factor of ${1}/{2}$ and may be expressed as
\be
\Gamma_{\rm ann}(\vec{x}_{S}) \, = \, \frac{1}{2} \, \sv \; n_{\chi}^{2}(\vec{x}_{S}) \, = \,
\frac{1}{2} \, \sv \left\{ {\displaystyle \frac{\rho_{\chi}(\vec{x}_{S})}{m_{\chi}}} \right\}^{2} .
\label{eq:ann_DM_conv}
\ee
The pre-factor would be $1/4$ in the case of Dirac/complex DM candidates. Two mediators $\phi$ are produced per DM annihilation. This leads to the mediator production rate at location $\vec{x}_{S}$
\be
q_{\phi}(\vec{x}_{S}) \, = \, 2 \times \Gamma_{\rm ann}(\vec{x}_{S}) \, = \,
\sv \left\{ {\displaystyle \frac{\rho_{\chi}(\odot)}{m_{\chi}}} \right\}^{2}
\left\{ {\displaystyle \frac{\rho_{\chi}(\vec{x}_{S})}{\rho_{\chi}(\odot)}} \right\}^{2} ,
\label{eq:q_phi}
\ee
where $\rho_{\chi}(\vec{x}_{S})$ and $\rho_{\chi}(\odot)$ respectively denote the DM density at point $\vec{x}_{S}$ and at the Sun.

In the Galactic frame where the DM species $\chi$ are at rest, mediators are monochromatic, with energy $E_{\phi} \equiv E_{\chi} \simeq m_{\chi}$. If the mediator lifetime in its rest frame is $\tau_\phi^0$, it is Lorentz boosted in the Galactic frame to $\tau_\phi = ({E_\phi}/{m_\phi})\,\tau_\phi^0 \simeq ({m_\chi}/{m_\phi})\,\tau_\phi^0$. Mediators move at a speed $v$ close to the celerity of light $c$ and decay in flight with the decay length $l_{\rm d} = v \, \tau_\phi$. Although the $\phi$ decay lifetime is dilated by a factor of ${\cal O}(10^{4})$ in the Galactic frame, the scenario which we explore here requires a significant amount of fine-tuning. We would like $\tau_\phi^0$ to be typically equal to ${\cal O}(3)$ years to get a decay length of 10\,kpc. In order not to ruin the successful primordial nucleosynthesis,  the  abundance of the long-lived mediator should be negligible after neutrino decoupling~\cite{Poulin:2015opa}.  This can be realized in particle models by, for instance, imposing strong self-annihilations among mediators at early times. 

The probability $P(>\!r)$ that a mediator propagates along a distance $r$ without decaying is just given by the factor $\exp(-r/l_{\rm d})$. Taking the derivative of $P(>\!r)$ with respect to $r$ yields the probability distribution function for a mediator to decay during propagation at distance $r$ from the production site $\vec{x}_{S}$
\be
P(r) \, = \, \frac{1}{l_{\rm d}} \, \exp \left( {- {r \over l_{\rm d}}} \right) \; .
\ee
As mediators are isotropically produced by DM annihilation, their flux decreases with distance as ${1}/{4 \, \pi \, r^{2}}$. The probability per unit volume that a mediator produced at $\vec{x}_{S}$ decays at position $\vec{x}$, yielding there a positron, may be expressed as
\be
\Gphi(\vec{x}_{S} \to \vec{x}) \, = \, {\displaystyle \frac{1}{4 \, \pi \, r^{2}}} \times P(r) \equiv
{\displaystyle \frac{e^{\displaystyle {-r}/{l_{\rm d}}}}{4 \, \pi \, l_{\rm d} \, r^{2}}} \;\; ,
\ee
where $r= |\vec{x}_S - \vec{x}|$. This function is the mediator propagator. It is normalized to unity when integrated on $d^{3}\vec{x}$.
The total production rate of positrons at point $\vec{x}$ is equal to the rate of mediator decays taking place there. It is given by the convolution
\be
q_{e^{\! +}}(\vec{x}) \, = \,
{\displaystyle \int} d^{3}\vec{x}_{S} \; q_{\phi}(\vec{x}_{S}) \; \Gphi(\vec{x}_{S} \to \vec{x}) \; .
\label{eq:q_pos_1}
\ee
If we are interested in the positron production rate per unit of volume and energy, we need to multiply the previous relation by the energy distribution ${dN_{e}}/{dE_{e}}$ yielded in the Galactic frame by each mediator decay to get
\be
q_{e^{\! +}}(\vec{x} , E_{e}) \, = \, {\displaystyle \frac{dN_{e}}{dE_{e}}} \;
{\displaystyle \int} d^{3}\vec{x}_{S} \; q_{\phi}(\vec{x}_{S}) \; \Gphi(\vec{x}_{S} \to \vec{x}) \; .
\label{eq:q_pos_2}
\ee
This relation can be generalized to any SM species produced by mediator decay.

Although the production rate of SM particles is quite different in the mediator scenario, it is still possible to get back to the conventional situation where SM products are directly injected where DM annihilates. Relations~(\ref{eq:q_phi}) and (\ref{eq:q_pos_2}) can be combined to recover the usual expression for SM production
\be
q_{\rm \, SM}(\vec{x} , E_{\rm \, SM}) \, = \,
\frac{1}{2} \, \sv \left\{ {\displaystyle \frac{\rho_{\rm eff}(\vec{x})}{m_{\chi}}} \right\}^{2} \,
{\displaystyle \frac{dN_{\rm SM}}{dE_{\rm \, SM}}} \; ,
\label{eq:ann_DM_eff}
\ee
as if each WIMP annihilation yielded a positron spectrum ${dN_{\rm SM}}/{dE_{\rm \, SM}}$ at the source equal to $2 \times {dN_{e}}/{dE_{e}}$ and DM were distributed with the effective density $\rho_{\rm eff}$ such that
\be
\rho_{\rm eff}^{2}(\vec{x}) \, = \,
{\displaystyle \int} d^{3}\vec{x}_{S} \; \rho_{\chi}^{2}(\vec{x}_{S}) \; \Gphi(\vec{x}_{S} \to \vec{x}) \; .
\label{eq:rho_eff}
\ee
According to this result also established by~\cite{Rothstein:2009pm}, the actual DM density $\rho_{\chi}$ is smeared by mediator propagation and attenuated on a scale of order the decay length $l_{\rm d}$. We expect dense DM spots such as the GC spike to be erased, all the more so if mediators are long-lived. On the contrary, regions where DM is not abundant could behave as if they contained much more DM than they actually do, should $\rho_{\rm eff}$ be exceedingly large with respect to $\rho_{\chi}$.

\begin{figure}[t!]
\centering
\includegraphics[width=0.49\textwidth]{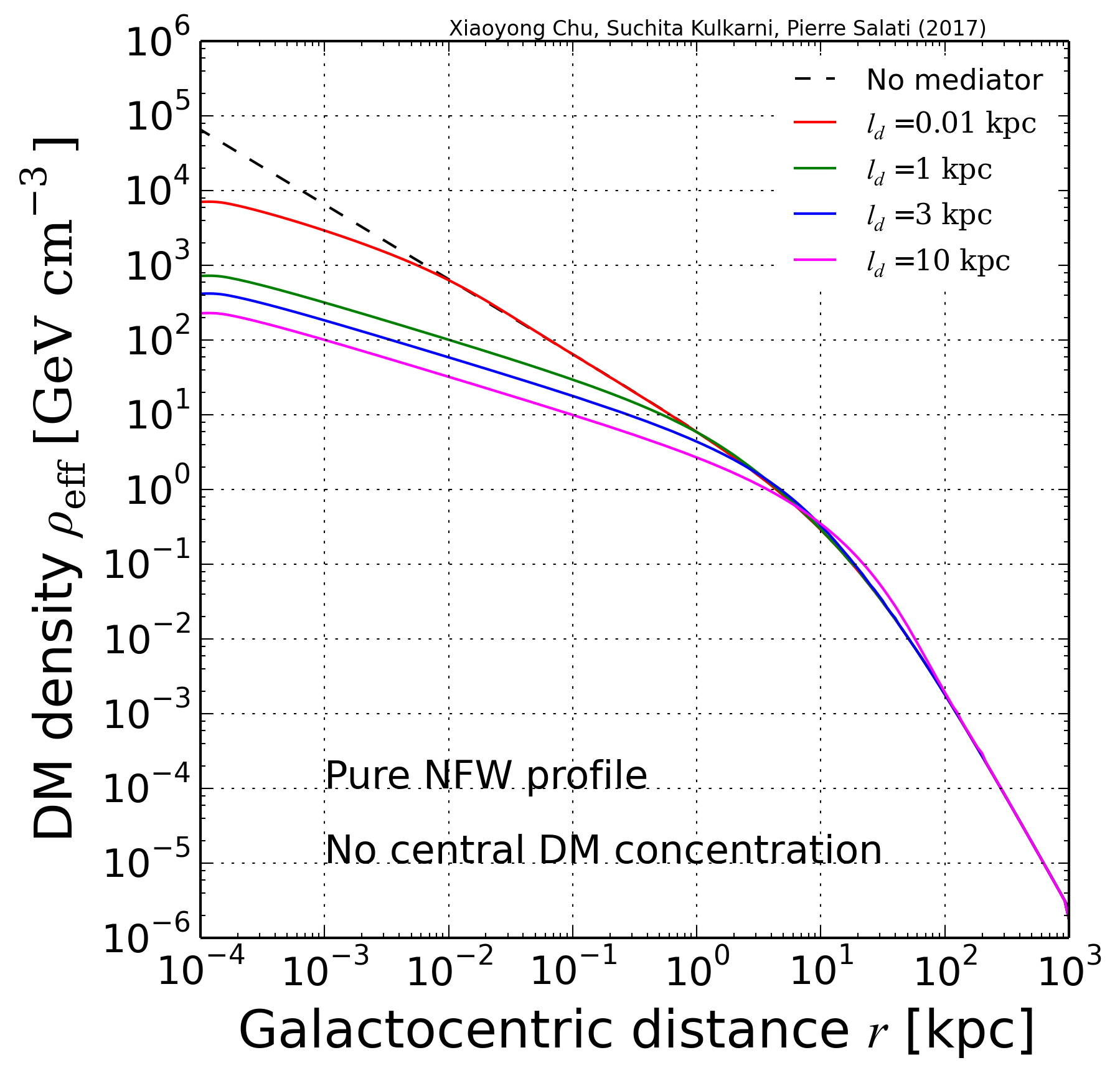}
\includegraphics[width=0.49\textwidth]{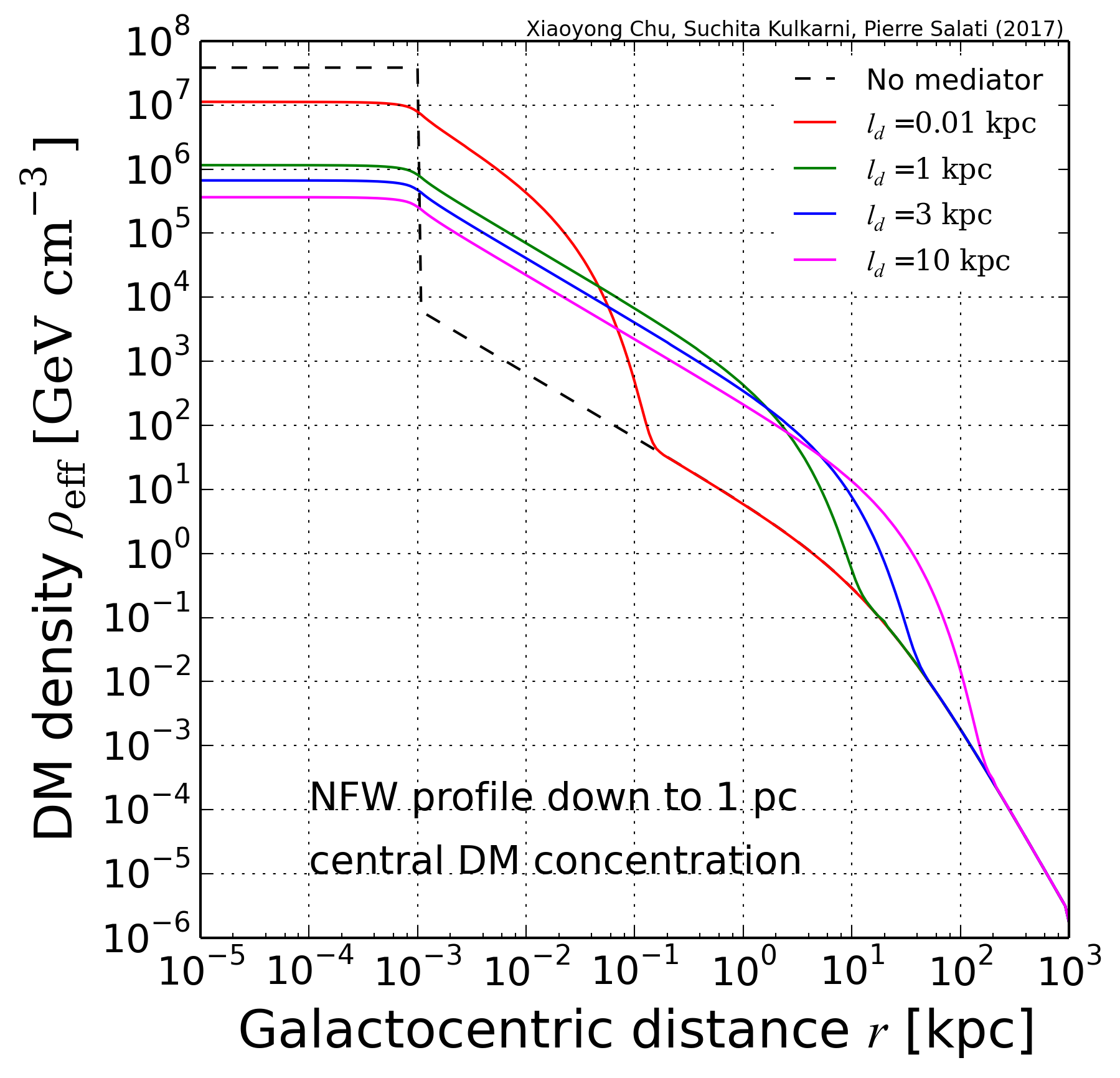}
\caption{
The mediator-smeared DM profile $\rho_{\rm eff}$ is plotted as a function of galactocentric radius $r$ for various decay length $l_{\rm d}$. In the left panel, the actual DM distribution $\rho_{\chi}$ is a pure NFW profile while in the right panel a very dense core has been added in the inner 1~pc. The long dashed black curves correspond to the unperturbed DM distribution $\rho_{\chi}$. The solid lines feature the effect of mediator smearing with a decay length $l_{\rm d}$ respectively equal to 0.01 (red), 1 (green), 3 (blue) and 10~kpc (magenta).}
\label{fig:medprofile}
\end{figure}

In the case of the Milky Way for which the DM distribution $\rho_{\chi}(r)$ is isotropic, relation~(\ref{eq:rho_eff}) translates into
\be
\rho_{\rm eff}^{2}(R) \, = \, {\displaystyle \int_{0}^{+ \infty}} \! {\displaystyle \frac{dl}{2 \, l_{\rm d}}} \;
{\displaystyle \int_{0}^{\pi}} \! d(- \cos\theta) \; \rho_{\chi}^{2}(r) \, \exp(-l/l_{\rm d})
\;\;\; \text{where} \;\;\;
r = \sqrt{R^{2} + l^{2} + 2 R \, l \cos{\theta}} \; .
\ee
In Fig.~\ref{fig:medprofile}, we have applied this expression to derive the mediator-smeared DM distribution for several mediator decay lengths $l_{\rm d}$. In the left panel, we have started from the pure  Navarro-Frenk-White (NFW) profile~\cite{Navarro:1995iw}
\be
\rho_{\chi}(r) \, = \, {\rho_0} \, \left\{ {\displaystyle \frac{r_{S}}{r}} \right\} \left\{ 1 + {\displaystyle \frac{r}{r_{S}}} \right\}^{-2} \;,
\label{eq:prof_NFW}
\ee
where the scale radius $r_S$ is taken to be 20~kpc. The normalization $\rho_0$ is such that the DM density in the solar neighbourhood is $\rho_{\chi}(\odot) = 0.4$ GeV/cm$^3$, at galactocentric distance $r_{\odot} = 8.2$ kpc.
In the right panel, a dense DM spot is added at the center of the previous NFW distribution. Inside a sphere of radius $r_0 = 1$~pc, the DM density is assumed to be homogeneous with a density larger than at its surface by a factor ${\cal N} = 5.9\times 10^3$. This number is borrowed from the recent proposal by~\cite{Kim:2017qaw} where long-lived mediators are used to explain the cosmic ray positron anomaly. If DM is densely packed at the GC, mediators boost the effective positron production rate at the Earth. The price to pay though is to assume the existence of the DM overdensity featured by the long-dashed curve in the right panel.
%

Mediator propagation and decays have two important effects on the DM distribution.
To commence, we note the attenuation of the central spike at small galactocentric distance $r$. The mediator-smeared density distribution $\rho_{\rm eff}$ differs significantly from the DM distribution $\rho_{\chi}$ in this region. In the left panel, the NFW profile which diverges as ${1}/{r}$ is smoothed to a shallower profile evolving as ${1}/{\sqrt{r}}$ for $r \lesssim l_d$, and the production of SM particles behaves as if DM decayed instead of annihilating. As already mentioned by~\cite{Rothstein:2009pm}, the effect of mediators is to make WIMP annihilations appear effectively as if they were decays. Notice that above a radius $r \sim l_{\rm d}$, the NFW profile is recovered. The smearing of mediators acts on a distance of order the decay length. In the limit where it is vanishingly small, mediators decay as soon as they are produced and there is no difference between the mediator scenario and the conventional situation for which SM species are injected where DM annihilates. When $l_{\rm d}$ is very small, the mediator propagator peaks at $\vec{x} = \vec{x}_{S}$ and both effective and actual DM distributions become equal. 
In the right panel, we also notice that the central density $\rho_{\rm eff}(r<r_{0})$ is significantly decreased with respect to the initial value $\rho_{\chi}$, up to two orders of magnitude for $l_{\rm d} = 10$~kpc.

The second and foremost effect of mediators is to redistribute DM outside the densest regions. This effect is conspicuous in the right panel where the DM effective density drops as ${1}/{r}$ outside the inner core and behaves as an enhanced NFW profile for galactocentric radii larger than 1~pc and smaller than the decay length. The DM density in the solar neighbourhood, for instance, is enhanced by a factor $\sim 45$ for $l_{\rm d} = 10$~kpc. If we assume that the effective DM density results only from the smearing of the central concentration lying inside the inner 1~pc, denoted as $\delta \rho_{\rm eff}$, a naive estimate of the boost is given by
\be
{\displaystyle \frac{\delta\rho_{\rm eff}(\odot)}{\rho_{\chi}(\odot)}} \, \simeq \, {\cal N} \, \left\{ 1 + {\displaystyle \frac{r_{\odot}}{r_{S}}} \right\}^{2}
\left\{ {\displaystyle \frac{r_{0}}{3 \, l_{\rm d}}} \right\}^{1/2} \! \exp(-{r_{\odot}}/{2 \, l_{\rm d}}) \sim 45 \;. \label{enhance:diffuse}
\ee
We have checked that given the NFW profile of DM density, $\delta\rho_{\rm eff}(\odot)$  serves as a very good approximation of the effective DM density $\rho_{\rm eff}$ at the Solar System. To summarize, the effective DM density is enhanced with respect to the actual one outside the dense regions where the opposite effect takes place. This is not surprising. According to definition~(\ref{eq:rho_eff}), the integrals over the entire space of both $\rho_{\chi}^{2}$ and $\rho_{\rm eff}^{2}$ must be equal. Mediators cannot deplete some regions from their DM contents without enhancing others. We finally note that at large galactocentric distances, i.e., when $r$ becomes much larger than $l_{\rm d}$, both profiles $\rho_{\chi}$ and $\rho_{\rm eff}$ are similar.

\subsection{The effect at the Solar System of adiabatic contraction at the GC}
\label{sec:blackhole}

The cosmic ray positron flux exhibits above a few GeV an excess with respect to the component named secondary produced by the interactions of high-energy protons and helium nuclei on interstellar gas. The anomaly was discovered by the PAMELA mission~\cite{Adriani:2008zr} and was recently confirmed with high accuracy by the AMS-02 collaboration~\cite{Aguilar:2013qda}. It has brought about a lot of excitement insofar as an excess in the antimatter cosmic ray spectra at the Earth is actually expected in the WIMP scenario.
The immediate problem though is the large boost the positron production rate requires to explain the excess compared to what is currently expected. High-energy positrons rapidly lose energy as they diffuse in the magnetic halo and those detected by PAMELA or AMS-02 must have been produced in the solar vicinity, where the DM density $\rho_{\chi}(\odot)$ is known to be of order $0.4$ GeV/cm$^3$. If the DM annihilation cross section is set equal to the thermal value of $3 \times 10^{-26}$ cm$^{3}$ s$^{-1}$, as required by the DM relic abundance, the positron production rate at the Earth is short by at least three orders of magnitude for a 1~TeV WIMP, hence the ongoing efforts of the community to enhance the local production rate of positrons while fulfilling a variety of astrophysical and cosmological requirements (for a review see for instance~\cite{Boudaud:2014dta}).

Recently, long-lived mediators have been proposed~\cite{Kim:2017qaw} to explain the positron anomaly. The effective DM density at the Earth can be significantly enhanced should a very dense DM concentration sits at the GC. Following~\cite{Kim:2017qaw}, if the DM density is enhanced there by a factor of ${\cal N} = 5.9 \times 10^{3}$ in the central 1~pc region, the boost in the local positron production rate reaches a value of $2 \times (45)^{2} \sim 4 \times 10^{3}$, in agreement with what is required, and the positron flux sourced by DM annihilation becomes compatible with the magnitude of the observations. The question naturally arises to determine if such a highly concentrated DM substructure can exist. DM is non dissipative and barely collapses. Numerical simulations of large scale structure formation indicate that DM can nevertheless condense at galactic centers, with distributions well described by NFW or Einasto profiles~\cite{Navarro:2008kc}. But even in the former case, the inner 1~pc at the center of the Milky Way would contribute to the effective DM density at the Earth a fraction ${\delta\rho_{\rm eff}(\odot)}/{\rho_{\chi}(\odot)} \sim 0.76\%$, i.e., several orders of magnitude below what is needed to account for the positron anomaly.

Given the difficulty with which DM collapses, the starting point of the Kim et al. proposal~\cite{Kim:2017qaw} is not a natural assumption. To the best of our knowledge, the only process that can lead to the desired DM density is adiabatic contraction. As proposed by~\cite{Gondolo:1999ef} and discussed in~\cite{Ullio:2001fb}, the formation of the massive black hole (BH) that sits at the GC could yield a very dense DM spike.
An initial DM sphere of radius $r_{i}$ contracts into a much smaller sphere with radius $r_{f}$ as baryons collapse at its center. DM does not interact with the infalling material but feels the deepening of the gravitational well as the central BH forms. The best conditions for the creation of a dense spike are met when (i) spherical symmetry holds and (ii) baryons condense slowly. The initial DM distribution is characterized by the NFW profile $\rho_{i}(r_{i}) \propto r_{i}^{- \gamma}$ with index $\gamma = 1$. The final distribution is also a power law with $\rho_{f}(r_{f}) \propto r_{f}^{- A}$. The conservation of the DM mass $M_{i}(r_{i}) = M_{f}(r_{f})$ during contraction translates into $r_{i}^{3 - \gamma} \propto r_{f}^{3 - A}$.
As the gravitational field is spherically symmetric, the orbital momenta of the DM particles are conserved. If all orbits are initially circular -- a strong assumption which we will relax below --  the slowness of the BH formation implies that they remain circular. A WIMP initially rotating at distance $r_{i}$ from the GC feels the DM mass $M_{i}(r_{i}) = M_{\rm DM}$. In the final state, it orbits at distance $r_{f}$ and feels the combined effect of the same DM mass $M_{f}(r_{f}) = M_{\rm DM}$ and of the BH mass $M_{\rm BH}$. The conservation of orbital momentum yields
\be
r_{i} \, M_{i}(r_{i}) \, = \, r_{f} \left\{ M_{\rm BH} + M_{f}(r_{f}) \right\} .
\label{eq:cons_L}
\ee
Very close to the GC, i.e., in the inner 65~pc in the case of the initial NFW profile~(\ref{eq:prof_NFW}), the DM mass $M_{\rm DM}$ is small compared to $M_{\rm BH}$ and adiabatic contraction has a strong effect on the DM density, resulting into $r_{f} \propto r_{i}^{4 - \gamma}$. From the conservation of both DM mass and WIMP orbital momentum, we readily infer that the final profile index goes as $A = {(9 - 2 \gamma)}/{(4 - \gamma)}$ and is equal to $7/3$ in the case of interest. We can go a step further and show that the final DM profile is given by
\be
\rho_{f}(r_{f}) \, = \, {\displaystyle \frac{\alpha^{2/3}}{3}} \, \rho_{\chi}(\odot) \,
\left\{ {\displaystyle \frac{r_{\odot}}{r_{f}}} \right\}^{7/3} ,
\ee
where the dimensionless parameter $\alpha$ is related to the BH mass through
\be
\alpha \, = \, {\displaystyle \frac{M_{\rm BH}}{2 \, \pi \, \rho_{\chi}(\odot) \, r_{\odot}^{3}}} \,
\left\{ 1 + {\displaystyle \frac{r_{\odot}}{r_{S}}} \right\} .
\ee
With $M_{\rm BH} = 4.6 \times 10^{6}$~M$_{\odot}$, we infer a DM mass of $2.95 \times 10^{5}$~M$_{\odot}$ in the inner 1~pc once the contraction has ended, with an initial value of $1.1 \times 10^{3}$~M$_{\odot}$. In the Kim et al. analysis~\cite{Kim:2017qaw}, the DM mass assumed to fill the same volume is a factor ${2 \, {\cal N}}/{3}$ larger than the initial value and amounts to $4.2 \times 10^{6}$~M$_{\odot}$.

In the case of adiabatic contraction, the DM mass at the GC is slightly smaller than the value claimed in~\cite{Kim:2017qaw}. But the DM distribution is spiky and not flat. The contribution of the central sphere with radius $r_{0}$ to the effective DM density in the solar neighbourhood, through long-lived mediators, may be expressed as
\be
\eta \, = \, \left\{ {\displaystyle \frac{\delta \rho_{\rm eff}(\odot)}{\rho_{\chi}(\odot)}} \right\}^{2} \! = \,
{\displaystyle \frac{1}{4 \, \pi \, r_{\odot}^{3}}} \times
\left\{ {\displaystyle \frac{r_{\odot}}{l_{\rm d}}} \; e^{\displaystyle - {r_{\odot}}/{l_{\rm d}}} \right\} \times \left\{
{\cal I} \equiv 
{\displaystyle \int_{0}^{r_{0}}} 4 \, \pi \, r_{f}^{2} \, dr_{f} \;
\left( {\displaystyle \frac{\rho_{f}(r_{f})}{\rho_{\chi}(\odot)}} \right)^{2} \right\}.
\label{eq_eta_1}
\ee
As the final profile $\rho_{f}$ scales as $r_{f}^{-7/3}$, the integral ${\cal I}$ is divergent at the center where the DM density becomes infinite. This is not physical. We remark that a WIMP population with an initial density $n_{\chi}^{0}$ annihilates with the density decreasing in time like
\be
n_{\chi}(t) \, = \,
{\displaystyle \frac{n_{\chi}^{0}}{1 \, + \, \sv \, n_{\chi}^{0} \, t}} \;.
\ee
We can revert the argument to set a maximal value of $n_{\rm ann} = {1}/{\sv \, t}$ to the WIMP density of a population that has been evolving since a time $t$. Assuming the age of the central BH is $\tau_{\rm BH}$, we find that the DM density cannot exceed a value of $\rho_{\rm ann} = {m_{\chi}}/{\sv \tau_{\rm BH}}$ inside the annihilation radius $r_{\rm ann} = \, r_{\odot} \, \alpha^{2/7} \, \beta^{3/7}$. The dimensionless parameter $\beta$ is defined as
\be
\beta \, = \, {\displaystyle \frac{\rho_{\chi}(\odot)}{m_{\chi}}} \, \sv \, {\displaystyle \frac{\tau_{\rm BH}}{3}} \;.
\ee
For a 1~TeV WIMP with thermal annihilation cross section, a BH age of 9 Gyr translates into an annihilation radius of $5.3 \times 10^{-3}$~pc within which the DM distribution exhibits a plateau with density $\rho_{\rm ann} = 1.17 \times 10^{11}$ GeV cm$^{-3}$.
Once adiabatic contraction and depletion through annihilation have taken place, the contribution of the central DM region to the effective local DM density $\rho_{\rm eff}(\odot)$ is
\be
\eta \, = \,
{\displaystyle \frac{14}{135}} \; \alpha^{6/7} \, \beta^{-5/7} \,
\left\{ {\displaystyle \frac{r_{\odot}}{l_{\rm d}}} \; e^{\displaystyle - {r_{\odot}}/{l_{\rm d}}} \right\} .
\ee
With a decay length $l_{\rm d} = 8.2$~kpc, for which the mediator effect at the Earth is maximal, we find a factor $\eta = 8 \times 10^{3}$ and a boost of the positron production rate in the solar neighbourhood reaching up to $2 \eta = 1.6 \times 10^{4}$, i.e., 4 times larger than in the Kim et al. scenario~\cite{Kim:2017qaw}.

This discussion may leave us with the impression that the formation of a massive BH at the GC can generate through adiabatic contraction a very dense DM spike, at least up to the level required to explain the positron excess. A word of caution is nevertheless mandatory. The scenario discussed above requires an idealized situation based on a few simplifying assumptions.
To commence, all orbits are assumed to be circular. This is a very crude hypothesis according to which the DM population close to the GC is infinitely cold, with radial velocities vanishing as $v(r) \propto \sqrt{r}$. But DM undergoes gravitational interactions with baryons and stars sitting there. Circular orbits should be replaced by an isotropic velocity-dispersion tensor. In this case, the DM phase space distribution in energy and orbital momentum can be related to the DM profile through Eddington's formula. Ullio et al.~\cite{Ullio:2001fb} have used it to show that a final spike still forms at the GC with the same radial profile $\propto {1}/{r^{7/3}}$ as found with the naive approach, but with a density twice as small. Replacing circular orbits by an isotropic velocity distribution yields an enhancement $\eta$ 4 times smaller than the previous estimate. The boost of the effective positron production rate in the solar vicinity falls down to $4 \times 10^{3}$, in marginal agreement with the value required by~\cite{Kim:2017qaw}.
Furthermore, we have assumed that the initial DM distribution follows a NFW profile all the way down to the GC. It is not clear whether or not results from numerical simulations can be extrapolated down to distances smaller than a few dozens of pc. Recent simulations tend to prefer the shallower Einasto profile and its central core~\cite{Dutton:2014xda}. Besides, baryons should also come into play and hinder the formation of an initial ${1}/{r}$ cusp. If a DM core stands initially at the GC with isotropic velocities, adiabatic contraction generates a $\rho_{f} \propto r^{-3/2}$ spike as shown by~\cite{Ullio:2001fb} and \cite{Capela:2014ita}. The positron boost factor is orders of magnitude smaller than the desired value.
Finally, the BH is assumed to grow (i) from a tiny initial mass (ii) slowly and (iii) at the very center of the initial NFW cusp. If the BH grows from an initial state containing already a substantial fraction of the final mass $M_{\rm BH}$, the collapse of the DM orbits is not as spectacular as if the growth started from nothing, insofar as what triggers the contraction is the difference between the initial and final masses in relation~(\ref{eq:cons_L}). Then, the BH formation may not be as slow as assumed. If the contraction were to be sudden, the initial NFW profile would lead to a $\rho_{f} \propto r^{-4/3}$ spike, very far from the required index of ${7}/{3}$. Finally, the BH seed may form off center the DM distribution and spiral to the center before accreting more material. The final spike would then be significantly smoother should the seed form with a mass at least equal to $1\%$ of the final value.

We conclude that in the best possible situation where a BH forms slowly at the exact center of an initial NFW cusp with isotropic velocities, the required value of $4 \times 10^{3}$ for the positron production rate enhancement is marginally obtained. Perturbing this scenario out of this idealized situation leads to a significant decrease of the boost and jeopardizes the claim by~\cite{Kim:2017qaw} that the positron excess can be explained with mediated DM.

\vspace{.1cm}
 
Besides, large enhancement of DM s-wave annihilation at present can also be obtained from the Sommerfeld effect~\cite{ArkaniHamed:2008qn}, if the light particle mediates DM self-interaction. Nevertheless, such an effect usually leads to even larger DM annihilation rates at recombination, which can hardly be made compatible with thermal freeze-out scenario~\cite{Bringmann:2016din}. Therefore, throughout this paper we assume a velocity-independent DM annihilation cross section and take  $3\times 10^{-26}$cm$^2$/s as its benchmark value. It is straightforward to re-scale our results to any other value of $\sv$ in the Galaxy at present.

\section{The flux of prompt particles and the gamma ray $J$-factor}
\label{sec:mediated_DM_ID}

Another important aspect of the signatures left by mediators is the direction toward which their decay products are emitted. This information is crucial for gamma rays since they propagate freely in space. For charged cosmic rays, whose momenta are isotropized by magnetic turbulences, we just need the total injection rate $q_{\rm \, SM}$, except in the case of the prompt species produced inside the ``last scattering'', or ``ballistic'' sphere.
We anticipate that the tools which we aim at building here combine information on (i) the flux of mediators along a primary direction $\vec{u}$ and (ii) the propensity with which their SM products move along the direction of interest  $\vec{w}$. We first discuss the mediator phase space distribution $\psdphi(\vec{x} , \vec{u})$ and then turn to the flux $\Phi_{\rm SM}(\vec{w} , E_{\rm \, SM})$ of the prompt particles which they produce.
We illustrate our discussion with photons but will apply our results to positrons in the next section.

\subsection{The mediator phase space distribution $\psdphi(\vec{x} , \vec{u})$}
\label{sec:PSD_mediator}

Mediators are mono-energetic and are produced by WIMP annihilations with momenta $\vec{p} \equiv p_{\phi} \, \vec{u}$, where $p_{\phi} \simeq m_{\chi}$ in the ultra-relativistic case. The mediator phase space distribution is the five dimensional function $\psdphi(\vec{x} , \vec{u})$ of position $\vec{x}$ and direction $\vec{u}$. It corresponds to the mediator flux $\Phi_{\phi} = v \, \psdphi$ where $v \simeq c$ is the velocity. 

To derive the mediator phase space distribution, let us consider the particles moving along the unit vector $\vec{u}$ up to the solid angle $d\Omega \equiv d^{2} \vec{u}$, and crossing the elementary surface $d\vec{S}$ during the time interval $dt$. They amount to
\be
d^{3}N_{\phi} \, = \, \left\{ \psdphi(\vec{x} , \vec{u}) \, d\Omega \right\} \times
\left\{ (\vec{v} \equiv v \, \vec{u}) \cdot d\vec{S} \right\} \times dt \;,
\ee
and originate from the cone with solid angle $d\Omega$ surrounding the opposite direction $- \vec{u}$. Without loss of generality, we may assume that the mediator velocity is perpendicular to the elementary surface $dS$. If so, the vectors $\vec{u}$ and $d\vec{S}$ are aligned and we get $d^{3}N_{\phi} = v \, \psdphi \, d\Omega \, dS \, dt$.
As steady state holds, the contribution to $\psdphi(\vec{x} , \vec{u})$ of the DM annihilations taking place inside a disk of thickness $dr$ located at distance $r$ from point $\vec{x}$ would be equal to
\be
d^{4}N_{\phi} \, = \, \left\{ r^{2} d\Omega \, dr \right\} \left\{ q_{\phi}(\vec{x}_{S}) \, dt \right\}
\left\{ {\displaystyle \frac{dS}{4 \, \pi \, r^{2}}} \right\} ,
\label{eq:psdpsi_1}
\ee
should mediators be stable. In the previous expression, the bracketed terms respectively refer to (i) the volume of the disk, (ii) the number of mediators produced per unit volume during the time interval $dt$, and (iii) the fraction of these reaching the surface $dS$ given that they are emitted isotropically. The source position is $\vec{x}_{S} = \vec{x} \, - \, r \, \vec{u}$.
Taking into account mediator decays and integrating over the distance $r$ leads to the phase space distribution and flux
\be
v \, \psdphi(\vec{x} , \vec{u}) \equiv \Phi_{\phi}(\vec{x} , \vec{u}) \, = \, {\displaystyle \frac{1}{4 \, \pi}} \;
{\displaystyle \int_{0}^{+\infty}} dr \; q_{\phi}(\vec{x}_{S}) \, e^{\displaystyle {-r}/{l_{\rm d}}} \;.
\label{eq:psdpsi_2}
\ee
This expression should be similar to the definition of the gamma ray flux yielded by canonical DM annihilation, and we recast it in terms of the mediator $J$-factor
\be
\psdphi(\vec{x} , \vec{u})\, = \, {\displaystyle \frac{1}{4 \, \pi \, v}} \,
{\displaystyle \frac{\sv}{m_{\chi}^{2}}} \, \left\{ J_{\phi}(- \vec{u}) \equiv
{\displaystyle \int_{0}^{+\infty}} dr \; \rho_{\chi}^{2}(\vec{x}_{S}) \, e^{\displaystyle {-r}/{l_{\rm d}}}
\right\} ,
\label{eq:psdpsi_3}
\ee
to which remote contributions are exponentially suppressed on a scale set by the decay length $l_{\rm d}$.

As an additional check, let us recover relation~(\ref{eq:q_pos_1}) from the expression which we have just established. For this, we assume as in section~\ref{sec:medprof} that each mediator decays to a single positron. At location $\vec{x}$, the number density of mediators is obtained by integrating $\psdphi$ over all possible directions
\be
n_{\phi}(\vec{x}) \, = \, {\displaystyle \int_{\displaystyle 4 \pi}} \psdphi(\vec{x} , \vec{u}) \; d^{2}\vec{u} \; .
\label{eq:density_phi}
\ee
The probability per unit time that a mediator decays is given by the inverse of its lifetime $\tau_{\phi}$. As the positron production rate is just the ratio ${n_{\phi}(\vec{x})}/{\tau_\phi}$, we get
\be
q_{e^{\! +}}(\vec{x}) \, = \,
{\displaystyle \frac{1}{\tau_\phi}} \; {\displaystyle \int_{\displaystyle 4 \pi}} \! d^{2}\vec{u} \times
{\displaystyle \frac{1}{4 \, \pi \, v}} \,
{\displaystyle \int_{0}^{+\infty}} dr \; q_{\phi}(\vec{x}_{S}) \; e^{\displaystyle {-r}/{l_{\rm d}}} \; .
\ee
The radial integral may be recast as
\be
q_{e^{\! +}}(\vec{x}) \, = \, {\displaystyle \frac{1}{4 \, \pi \, v \, \tau_\phi}} \;
{\displaystyle \int_{\displaystyle 4 \pi}} d^{2}\vec{u} \; {\displaystyle \int_{0}^{+\infty}} r^{2} \, dr \times q_{\phi}(\vec{x}_{S}) \times
{\displaystyle \frac{e^{\displaystyle {-r}/{l_{\rm d}}}}{r^{2}}} \; .
\ee
Noticing that $d\Omega \equiv d^{2}\vec{u}$, we can identify the volume element $d^{3}\vec{x}_{S}$ with $r^{2} d\Omega \, dr$ and translate the last expression into
\be
q_{e^{\! +}}(\vec{x}) \, = \,
{\displaystyle \int} d^{3}\vec{x}_{S} \; q_{\phi}(\vec{x}_{S}) \;
{\displaystyle \frac{e^{\displaystyle {-r}/{l_{\rm d}}}}{4 \, \pi \, l_{\rm d} \, r^{2}}} \; .
\label{eq:q_pos_3}
\ee
We have just obtained the convolution~(\ref{eq:q_pos_1}) between the mediator production rate $q_{\phi}$ at position $\vec{x}_{S}$ with the mediator propagator where $r = |\vec x_S -\vec x|$ and $l_{\rm d} = v \, \tau_\phi$.

\subsection{The flux of prompt particles produced by mediator decay}
\label{sec:prompt_flux}

We focus here on the flux, at the position $\vec{x}_{\odot}$ of the Earth, of the gamma rays produced by the mediators decaying in flight at location $\vec{x}$ and created by DM annihilation at the source $\vec{x}_{S}$. Mediators propagate freely from $\vec{x}_{S}$ to $\vec{x}$. So do photons from $\vec{x}$ to $\vec{x}_{\odot}$. 
In the Galactic frame, each mediator yields the energy and angular distribution of gamma rays $\Dgam(\vec{u} , \vec{w} , E_{\gamma})$, with $E_{\gamma}$ the photon energy. The unit vectors $\vec{u}$ and $\vec{w}$ are respectively aligned with the mediator and photon momenta, i.e., $\vec{p}_{\phi} = p_{\phi} \, \vec{u}$ while $\vec{p}_{\gamma} = E_{\gamma} \, \vec{w}$.
Should mediators be scalars, a somewhat reasonable assumption, photons would be produced isotropically in the mediator rest frame. A boost to the Galactic frame would result into an axisymmetric distribution around the initial direction of motion $\vec{u}$. The only angular dependence of $\Dgam$ would be on the photon polar angle $\theta$. More complicated situations are in principle possible if, for instance, mediators are vector particles polarized at the source by DM annihilation. We will keep then the notations as general as possible, noticing though that the angular information is just encoded in the difference $\vec{w} - \vec{u}$.
Integrating $\Dgam$ over the final direction of motion $\vec{w}$ yields the photon energy distribution per mediator decay
\be
{\displaystyle \frac{dN_{\gamma}}{dE_{\gamma}}} \, = \,
{\displaystyle \int_{\displaystyle 4 \pi}} \Dgam(\vec{w} - \vec{u} , E_{\gamma}) \; d^{2}\vec{w} \; .
\label{eq:gam_spectrum_1}
\ee
We note in passing that since $\vec{u}$ and $\vec{w}$ enter the distribution $\Dgam$ through their difference, we can just as well set the photon momentum direction $\vec{w}$ constant and integrate over the mediator momentum direction $\vec{u}$ to get
\be
{\displaystyle \frac{dN_{\gamma}}{dE_{\gamma}}} \, = \,
{\displaystyle \int_{\displaystyle 4 \pi}} \Dgam(\vec{w} - \vec{u} , E_{\gamma}) \; d^{2}\vec{u} \; .
\label{eq:gam_spectrum_2}
\ee
A few examples of the distribution $\Dgam$ are given in appendix.~\ref{app:positron},  where the kinematics of two-body and three-body mediator decays are studied.

Let us consider now the photons at the Earth (i) whose momenta are aligned with $\vec{w}$ while pointing inside the solid angle $d\Omega$ and (ii) with energy $E_{\gamma}$ up to $dE_{\gamma}$. The amount that crosses the elementary surface $dS$ during the time interval $dt$ defines the flux
\be
d^{4}N_{\gamma} \, = \, \Phi_{\gamma}(\vec{w} , E_{\gamma}) \, d\Omega \, dE_{\gamma} \, dS \, dt \;,
\label{eq:gam_flux_1}
\ee
which we aim at.
Among that population, the number of photons that are produced by mediators whose momenta are aligned with $\vec{u}$ up to $d\Omega_{\vec{u}} \equiv d^{2}\vec{u}$, and which decay at distance $r'$ up to $dr'$ from the Earth, is given by the product
\be
d^{6}N_{\gamma} \, = \, \left\{ r'^{2} d\Omega \, dr' \right\}
\left\{ {\displaystyle \frac{dt}{\tau_\phi}} \times \psdphi(\vec{x} , \vec{u}) \, d\Omega_{\vec{u}} \right\}
\left\{  \Dgam(\vec{w} - \vec{u} , E_{\gamma}) \times dE_{\gamma} \times {\displaystyle \frac{dS}{r'^{2}}} \right\} .
\label{eq:gam_flux_2}
\ee
The meaning of the bracketed terms is more involved than for expression~(\ref{eq:psdpsi_1}). They respectively describe (i) the volume of the disk, seen through the solid angle $d\Omega$ at distance $r'$ up to $dr'$, inside which mediators are considered, (ii) the number density of these with momenta directed along $\vec{u}$ up to $d\Omega_{\vec{u}}$ which decay during the time interval $dt$, and (iii) the number of photons emitted per mediator decay, with energy $E_{\gamma}$ up to $dE_{\gamma}$, whose directions point along $\vec{w}$ and cross the perpendicular detecting surface $dS$, hence the photon solid angle $d\Omega_{\vec{w}} \equiv d^{2}\vec{w} = {dS}/{r'^{2}}$. The mediators decay at position $\vec{x} = \vec{x}_{\odot} \, - \, r' \vec{w}$.
Tossing expressions~(\ref{eq:gam_flux_1}) and (\ref{eq:gam_flux_2}) together yields the general expression for the gamma ray flux
\be
\Phi_{\gamma}(\vec{w} , E_{\gamma}) \, = \,
{\displaystyle \frac{1}{\tau_\phi}} \;
{\displaystyle \int_{0}^{\displaystyle r_{\rm max}}} dr' \;
{\displaystyle \int_{\displaystyle 4 \pi}} \! d^{2}\vec{u} \times \psdphi(\vec{x} , \vec{u}) \times \Dgam(\vec{w} - \vec{u} , E_{\gamma}) \; .
\label{eq:gam_flux_3}
\ee
The mediator phase space distribution $\psdphi$ is given by equation~(\ref{eq:psdpsi_2}). For photons, the line of sight (los) integral runs from $0$ to $r_{\rm max}$ infinite. This will not be the case for prompt positrons in section~\ref{sec:anisotropy}. As mentioned in~\cite{Rothstein:2009pm}, a few limiting cases can be readily outlined:

(i) Let us first assume that mediators decay instantaneously, with vanishing decay length $l_{\rm d}$. The mediator phase space distribution at location $\vec{x}$ simplifies into
\be
\psdphi(\vec{x} , \vec{u}) \, = \, {\displaystyle \frac{\tau_{\phi}}{4 \, \pi}} \; q_{\phi}(\vec{x}) \;,
\ee
as the radial integral~(\ref{eq:psdpsi_2}) picks up the mediator production rate at position $\vec{x}_{S} = \vec{x}$. Mediator velocities are now isotropic. Inserting this expression into equation~(\ref{eq:gam_flux_3}) and using the normalization condition~(\ref{eq:gam_spectrum_2}) leads to the relation
\be
\Phi_{\gamma}(\vec{w} , E_{\gamma}) \, = \, {\displaystyle \frac{1}{4 \, \pi}} \,
{\displaystyle \frac{dN_{\gamma}}{dE_{\gamma}}} \;
{\displaystyle \int_{0}^{+ \infty}} ds \; q_{\phi} \! \left\{ \vec{x}(s) = \vec{x}_{\odot} \, - \, s \, \vec{w} \right\} ,
\ee
which can be recast into the canonical form
\be
\Phi_{\gamma}(\vec{w} , E_{\gamma}) \, = \, \left\{ {\displaystyle \frac{1}{4 \, \pi}} \,
{\displaystyle \frac{\sv}{m_{\chi}^{2}}} \, {\displaystyle \frac{dN_{\gamma}}{dE_{\gamma}}} \right\} \times
\left\{ J(- \vec{w}) \equiv {\displaystyle \int_{0}^{+\infty}} ds \; \rho_{\chi}^{2} \! \left\{ \vec{x}(s) \right\} \right\} .
\label{eq:gam_flux_trad_1}
\ee
The flux is the product of a particle physics part with the usual astrophysical $J$-factor. Remember that 2 mediators are produced per DM annihilation, hence a difference of a factor $1/2$ with respect to the gamma ray flux from Majorana DM.

 (ii) Another limiting situation is realized when mediators are ultra-relativistic, which is the case in most configurations. In this regime, SM decay products are boosted in the forward direction, with their momenta aligned with the mediator momentum. The vectors $\vec{u}$ and $\vec{w}$ are very close to each other. We may simplify the photon distribution $\Dgam$ into
\be
\Dgam(\vec{w} - \vec{u} , E_{\gamma}) \, \simeq \, {\displaystyle \frac{dN_{\gamma}}{dE_{\gamma}}} \times
\delta^{2}(\vec{w} - \vec{u}) \;.
\ee
Inserting it into the general flux expression~(\ref{eq:gam_flux_3}) and developing the mediator phase space distribution $\psdphi(\vec{x} , \vec{w})$ with the help of equation~(\ref{eq:psdpsi_3}) lead to the double integral
\be
\Phi_{\gamma}(\vec{w} , E_{\gamma}) \, = \, {\displaystyle \frac{1}{4 \, \pi}} \,
{\displaystyle \frac{\sv}{m_{\chi}^{2}}} \, {\displaystyle \frac{dN_{\gamma}}{dE_{\gamma}}} \times
{\displaystyle \int_{0}^{+\infty}} dr'  {\displaystyle \int_{0}^{+\infty}} dr \;
\rho_{\chi}^{2} \! \left\{ \vec{x}_{S} = \vec{x}_{\odot} - (r + r') \vec{w} \right\} \;
{\displaystyle \frac{e^{\displaystyle {-r}/{l_{\rm d}}}}{l_{\rm d}}} \;,
\label{eq:J_factor_UR_mediator_0}
\ee
where we must keep in mind that the collimated production of SM species makes $\vec{u}$ and $\vec{w}$ equal. With the new variables $r$ and $s = r + r'$, the last expression boils down to relation~(\ref{eq:gam_flux_trad_1}) where the astrophysical $J$-factor is now given by
\be
J_{\rm eff}(- \vec{w}) \, = \,
{\displaystyle \int_{0}^{+\infty}} ds \; \rho_{\chi}^{2} \! \left\{ \vec{x}_{S}(s) \right\}
\left\{1 - e^{\displaystyle {-s}/{l_{\rm d}}} \right\} .
\label{eq:J_factor_UR_mediator}
\ee
The only difference with the canonical case is the presence of the term $1 - P(>\!s)$ in the los integral. Mediators are isotropically produced by DM annihilations at position $\vec{x}_{S}$. Among those moving to the Earth, a fraction $P(>\!s)$ makes it while the rest is converted into signal photons. The contribution of the nearby region to the $J$-factor is suppressed on a scale of order the decay length. As a check, notice that we recover the conventional value of $J$ when $l_{\rm d}$ is vanishingly small, as already showed.

(iii) The results of~\cite{Rothstein:2009pm} are based on the restrictive assumption that mediators are non-relativistic. This is possible when the mediator mass $m_{\phi}$ is close to the WIMP mass $m_{\chi}$, but requires fine-tuning. In this limit, mediators decay isotropically in the Galactic frame and are distributed according to
\be
\Dgam(\vec{w} - \vec{u} , E_{\gamma}) \, \simeq \,
{\displaystyle \frac{dN_{\gamma}}{dE_{\gamma}}} \times {\displaystyle \frac{1}{4 \, \pi}} \;.
\ee
This leads to the flux
\be
\Phi_{\gamma}(\vec{w} , E_{\gamma}) \, = \,
{\displaystyle \frac{1}{4 \, \pi \, \tau_\phi}} \; {\displaystyle \frac{dN_{\gamma}}{dE_{\gamma}}} \,
{\displaystyle \int_{0}^{+\infty}} ds \; n_{\phi} \! \left\{ \vec{x}(s) = \vec{x}_{\odot} \, - \, s \, \vec{w} \right\} ,
\ee
and to the $J$-factor
\be
J_{\rm eff}(- \vec{w}) \, = \,
{\displaystyle \int_{0}^{+\infty}} ds \; \rho_{\rm eff}^{2} \! \left\{ \vec{x}(s) \right\} ,
\label{eq:J_factor_NR_mediator}
\ee
where the smeared DM density $\rho_{\rm eff}$ comes now into play. From Fig.~\ref{fig:medprofile} and our discussion of section~\ref{sec:eff_DM_rho}, we anticipate a decrease of $J_{\rm eff}$ at the centers of dense systems counterbalanced by an increase in their outskirts, compared to the canonical case.

\subsection{The gamma ray $J$-factor in the presence of mediator Dark Matter}
\label{sec:J_factor}

These considerations on the gamma ray flux yielded at the Earth by mediated DM lead us to define the generalized or effective $J$-factor through
\be
\Phi_{\gamma}(\vec{w} , E_{\gamma}) \, = \, {\displaystyle \frac{1}{4 \, \pi}} \,
{\displaystyle \frac{\sv}{m_{\chi}^{2}}} \, {\displaystyle \frac{dN_{\gamma}}{dE_{\gamma}}} \times J_{\rm eff}(- \vec{w}) \;,
\label{eq:gam_flux_trad_2}
\ee
where the new term
\be
J_{\rm eff}(- \vec{w}) \, = \,
\left\{ {\displaystyle \frac{dN_{\gamma}}{dE_{\gamma}}} \right\}^{-1}
{\displaystyle \int_{\displaystyle 4 \pi}} \! d^{2}\vec{u} \,
{\displaystyle \int_{0}^{+\infty}} \! dr  {\displaystyle \int_{0}^{+\infty}} \! dr' \;
\rho_{\chi}^{2} \! \left\{ \vec{x}_{S}(\vec{u},r,r') \right\} \, \Dgam(\vec{w} - \vec{u} , E_{\gamma})  \;
{\displaystyle \frac{e^{\displaystyle {-r}/{l_{\rm d}}}}{l_{\rm d}}}
\label{eq:J_eff_1}
\ee
contains all pertinent information on how mediators propagate and decay. The location of the source is at $\vec{x}_{S} = \vec{x}_{\odot} \, - \, r' \vec{w} \, - \, r \, \vec{u}$. Once the photon direction $\vec{w}$ is set, $ J_{\rm eff}$ is an intricate function of the mediator direction $\vec{u}$ and of the distances $r$ and $r'$. In the previous section, we have analyzed limiting situations where the $J$-factor can be easily calculated along the los.

We would like now to extend this discussion to the general case, going a step further than~\cite{Rothstein:2009pm}. Given a Galactic DM distribution $\rho_{\chi}$ and photons moving toward the direction\footnote{This corresponds to a los pointing in the opposite direction $- \vec{w}$.} set by $\vec{w}$ , we anticipate that the mediated to conventional $J$-factor ratio ${J_{\rm eff}}/{J}$ should depend on (i) the photon energy to DM mass ratio ${E_{\gamma}}/{m_{\chi}}$, (ii) the mediator velocity $v = \beta_{\phi} \, c$ and (iii) the decay length $l_{\rm d}$.
For illustration purposes, we will consider hereafter the case of a scalar mediator decaying into two photons. This is not an unrealistic possibility. Nature provides us with examples such as the neutral pion for which this is the main decay channel. In the recent past, when rumor had it that a two-photon resonance with mass 750\,GeV was on the verge of being found at the LHC, a wealth of theoretical proposals flourished to make it a quite plausible solution.

In the mediator rest frame, photons are emitted isotropically and are mono-energetic, with energy $E^{\star}_{\gamma} \equiv {m_{\phi}}/{2}$. In the Galactic frame, they receive a Lorentz boost from the mediator momentum, and thus have an energy $E_{\gamma}$ related to their polar angle $\theta$ through
\be
E_{\gamma} \, = \, {\displaystyle \frac{{m_{\phi}^{2}}/{2 m_{\chi}}}{(1 - \beta_\phi \cos\theta)}} \;.
\label{eq:E_gam_cos_polar}
\ee
The polar angle is defined between the mediator and photon directions $\vec{u}$ and $\vec{w}$, while the mediator velocity is $\beta_{\phi} = (1 - {m_{\phi}^{2}}/{m_{\chi}^{2}})^{1/2}$. Another consequence of the isotropic distribution of photons in the mediator rest frame is their ``box-shaped'' spectrum~\cite{Ibarra:2012dw} in the Galactic frame where
\be
{\displaystyle \frac{dN_{\gamma}}{dE_{\gamma}}} \, = \, {\displaystyle \frac{2}{\beta_{\phi} m_{\chi}}}
\;\;\; \text{for}\;\;\;
\left| E_{\gamma} - {{m_{\chi}}\over{2}} \right| \le {\beta_{\phi}m_{\chi}\over{2}} \;.
\ee
The derivation of the energy and angular distribution $\Dgam$ is detailed in appendix~\ref{app:positron}. We convert it here into
\be
\Dgam(\cos\theta , E_{\gamma}) \, = \, {\displaystyle \frac{2}{\beta_{\phi} m_{\chi}}} \times
{\displaystyle \frac{\delta \left( \cos\theta - \cos\theta_{0} \right)}{2 \, \pi}} \;,
\ee
where the angle $\theta_{0}$ fulfils condition~(\ref{eq:E_gam_cos_polar}). The expression of the effective $J$-factor simplifies into the triple integral
\be
J_{\rm eff}(- \vec{w}) \, = \,
{\displaystyle \int_{0}^{+\infty}} \! dr' \;
{\displaystyle \int_{0}^{2 \pi}} \, {\displaystyle \frac{d\varphi}{2 \, \pi}} \;
{\displaystyle \int_{0}^{+\infty}} \! dr \times
\rho_{\chi}^{2} \! \left\{ \vec{x}_{S}(\varphi,r,r') \right\} \times
{\displaystyle \frac{e^{\displaystyle {-r}/{l_{\rm d}}}}{l_{\rm d}}} \;.
\label{eq:J_eff_2}
\ee
A first integral is performed along the los, in the direction from which photons originate. Once a point $D$ is selected on the los, a second integral is carried out on the azimuthal angle $\varphi$ which defines the mediator direction of motion $\vec{u}$ with respect to $\vec{w}$. The polar angle $\theta_{0} \equiv (\vec{w} , \vec{u})$ is set by the photon energy $E_{\gamma}$ through relation~(\ref{eq:E_gam_cos_polar}). The final integral is performed over the distance $r$ that separates the source $S$, where mediators are produced by DM annihilation, from the position $D$, where they decay and yield photons which are collected at the Earth.
Thus, for each point $D$, a sum is carried out over the sources $S$ that belong to the cone originating from it and opening around the los with polar angle $\theta_{0}$.

\begin{figure}[t!]
\centering
\includegraphics[width=0.49\textwidth]{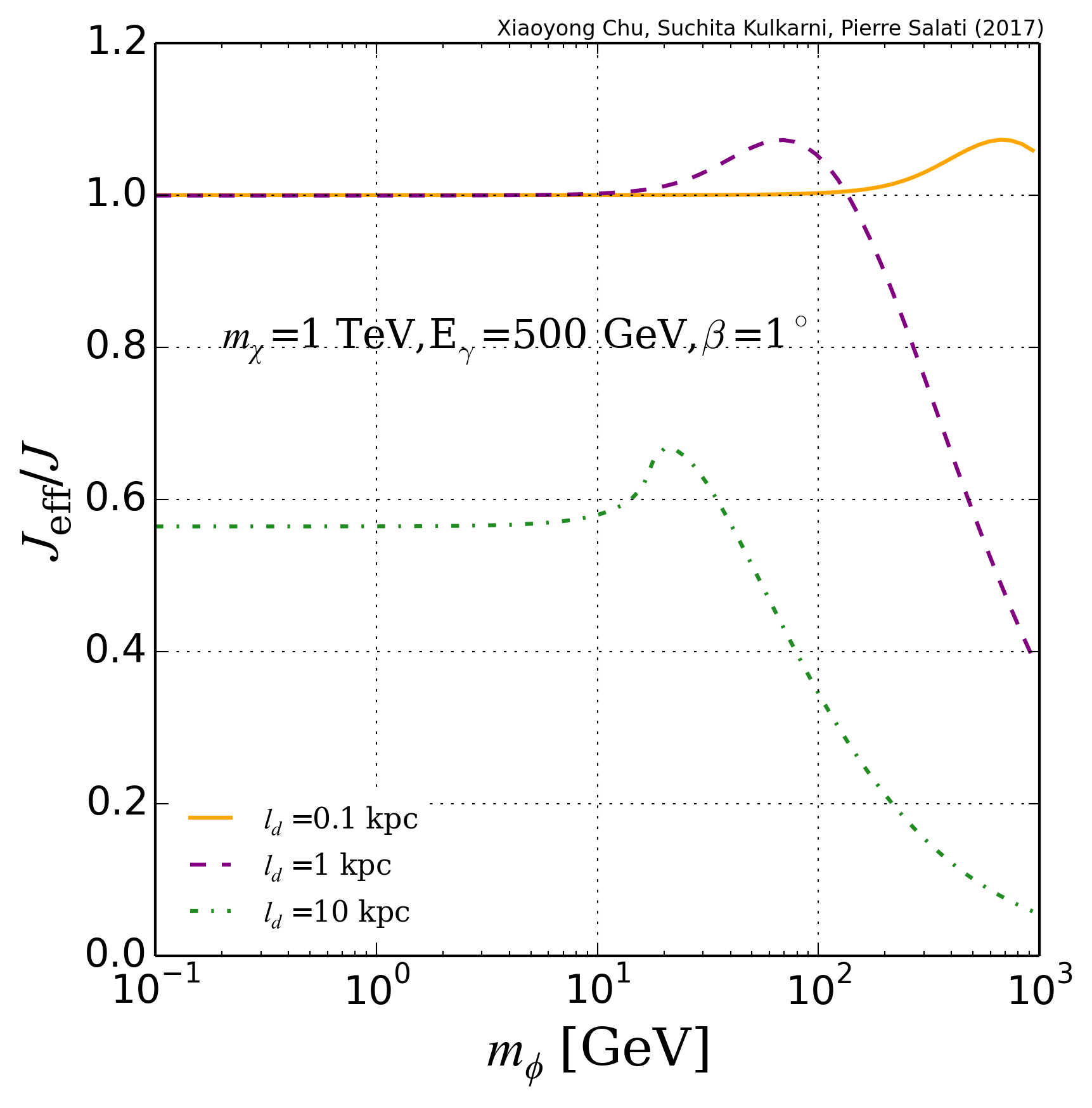}
\includegraphics[width=0.49\textwidth]{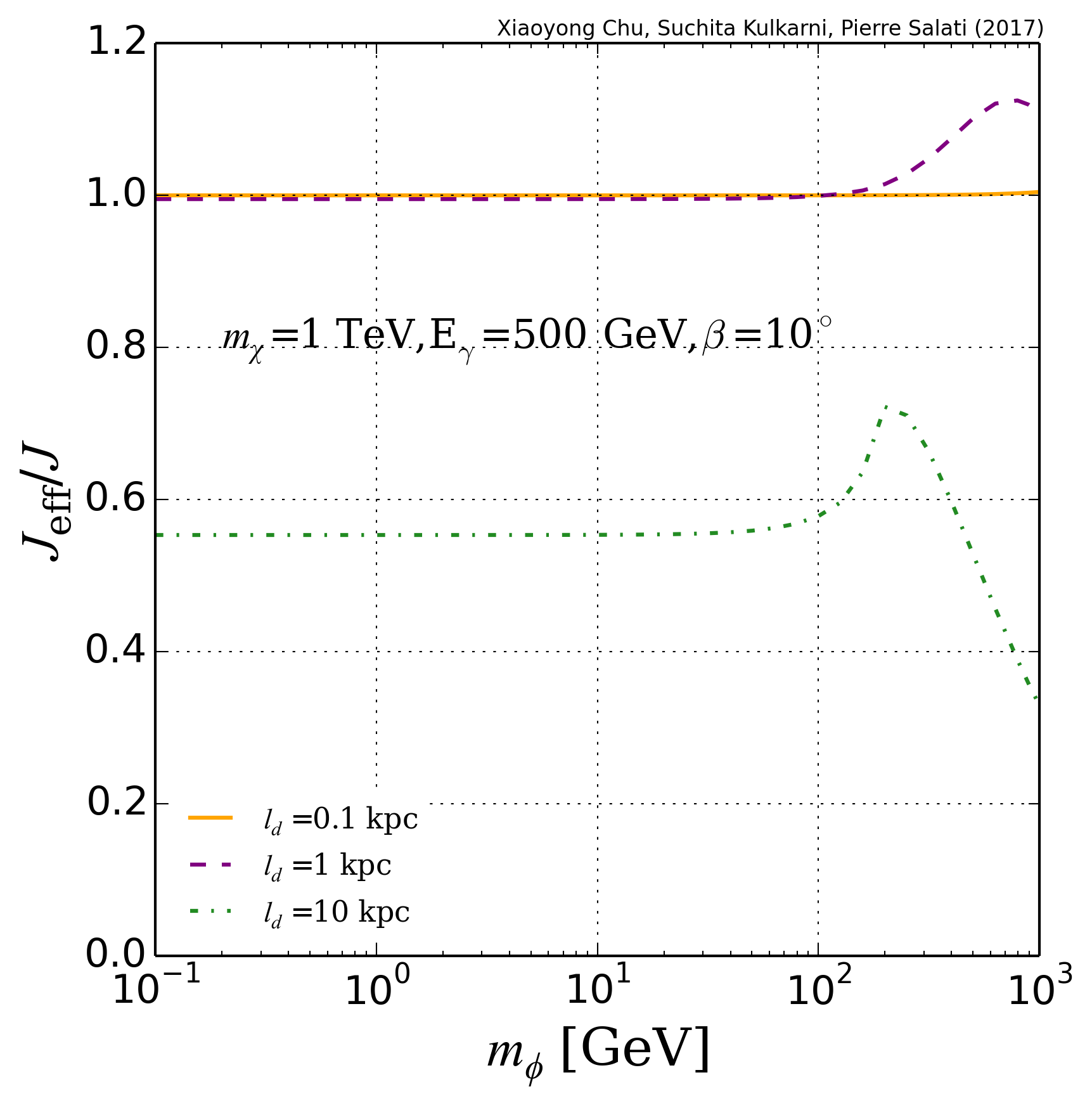}
\caption{
Dependence of the effective $J$-factor on mediator mass and decay length for two observing angles:  $1^{\circ}$ and $10^{\circ}$.  The WIMP mass is taken to be 1\,TeV and the observed gamma ray energy is $500$\,GeV.}
\label{Jfctr:smear_m_phi}
\end{figure}

We have numerically estimated $J_{\rm eff}$ for a DM mass of 1\,TeV and a photon energy of 500\,GeV. In the Galactic frame, the value $E_{\gamma} = {m_{\chi}}/{2}$ falls in the middle of the photon spectrum, irrespective of the mediator velocity $\beta_{\phi}$. Moreover, with that choice, the opening angle $\theta_{0}$ of the cones is such that $\cos\theta_{0} = \beta_{\phi}$, or alternatively $\sin\theta_{0} = {m_{\phi}}/{m_{\chi}}$.
In Fig.~\ref{Jfctr:smear_m_phi}, the ratio of the $J$-factor to its NFW expectation is plotted as a function of mediator mass $m_{\phi}$, for three values of the decay length $l_{\rm d}$. In the left and right panels, the angle $\beta$ between the los and the GC is respectively set equal to $1^{\circ}$ and $10^{\circ}$. The DM distribution is given by the NFW profile~(\ref{eq:prof_NFW}).
We first remark that all curves are flat over a significant portion of the mediator mass range, starting from $0.1$\,GeV upward. As the mediator mass $m_{\phi}$ becomes small with respect to the DM mass $m_{\chi}$, the angle $\theta_{0} \propto {m_{\phi}}/{m_{\chi}}$ decreases and the cones over which the DM density is integrated close. Photons are produced along the same direction as their decaying progenitors. As discussed in section~\ref{sec:prompt_flux}, $J_{\rm eff}$ is in this regime well approximated by relation~(\ref{eq:J_factor_UR_mediator}) and no longer depends on the mediator and DM masses. The decay length alone comes into play, with a result all the more suppressed as $l_{\rm d}$ is large. For $0.1$ and $1$\,kpc, the $J$-factor is equal to its NFW expectation while it amounts only to 57\% of it for $10$\,kpc. In that case, mediators come from regions located far away from the los, where the DM density is on average smaller.
We also notice that when $l_{\rm d}$ is small, the ${J_{\rm eff}}/{J}$ ratio is equal to 1 whatever the mediator mass. This is clear for the solid yellow line in the right panel. In that case, mediators decay as soon as they are produced, as explained in section~\ref{sec:prompt_flux}. The $J$-factor becomes sensitive only to the DM distribution on the los and not around it.
To summarize, the behaviour of the solid yellow and dashed purple curves of Fig.~\ref{Jfctr:smear_m_phi} can be understood as the combined consequence of small mediator masses and short decay lengths.

The curves also exhibit bumps at particular locations, with a $J$-factor larger than its NFW expectation.
This is the case of the solid yellow curve in the left panel for mediator masses close to the DM mass. In this configuration where the observing angle $\beta = 1^{\circ}$, the distance of closest approach of the los to the GC is $a = r_{\odot} \sin\beta = 0.14$\,kpc, to be compared to the decay length $l_{\rm d} = 0.1$\,kpc. The DM density is very large at the GC where an intense flux of mediators is emitted. A fraction $\exp(-{a}/{l_{\rm d}}) \sim$ 25\% of the particles makes it to the los. If the photon to mediator angle $\theta_{0}$ is close to $90^{\circ}$, which is the case when $m_{\phi}$ is near $m_{\chi}$, the decay photons are produced along the los toward the Earth, hence a bump. The same explanation holds for the dashed purple curve of the right panel where $a = 1.4$\,kpc whilst $l_{\rm d} = 1$\,kpc.

A generic explanation for the bumps is in order at this stage. There are geometric configurations where some of the cones over which integral~(\ref{eq:J_eff_2}) is calculated cross the GC region, allowing mediators produced there to reach the los, and deliver photons which propagate along it toward the Earth. These configurations correspond to particular combinations of the observing and aperture angles $\beta$ and $\theta_{0}$.
To commence, let us consider the case where the decay length $l_{\rm d}$ is larger than the galactocentric distance $r_{\odot}$. A substantial fraction of the mediators from the GC reach the los. In order for the photons which they yield to be emitted along the los, the aperture angle $\theta_{0}$ must be larger than the observing angle $\beta$. We expect a contribution from the GC to the $J$-factor to appear as soon as $\theta_{0}$ overreaches $\beta$. This translates into $m_{\phi}$ larger than $\sin\beta \times m_{\chi}$. Since the DM concentration at the GC has a finite extension, the maximum is actually reached for a value of $\theta_{0}$ slightly larger than $\beta$. For the dashed-dotted green curves of Fig.~\ref{Jfctr:smear_m_phi}, a bump develops above 17\,GeV (left panel) and 170\,GeV (right panel). The maxima are respectively reached for $m_{\phi} = 20$ and 200\,GeV.

In the opposite situation of a decay length smaller than the galactocentric distance, an additional constraint appears. As above, the aperture angle $\theta_{0}$ must be larger than the observing angle $\beta$. When this condition is met, the points of the los where mediators from the GC can inject photons toward the Earth span a distance ${\Delta s} \simeq {\Delta r}/{\sin\theta_{0}}$, where ${\Delta r}$ is the GC radial extension. But now the contribution of the GC to the $J$-factor scales as $\exp(-{r}/{l_{\rm d}})$, with $r$ the distance between the los and the GC. This term can be significantly suppressed if the decay length is small. The distance $r$ is related to the angles $\beta$ and $\theta_{0}$ through the triangular identity
\be
a \, = \, r \, \sin\theta_{0} \, = \, r_{\odot} \, \sin \beta \;,
\ee
where $a$ is the distance of closest approach of the los to the GC. According to relation~(\ref{eq:J_eff_2}), the contribution of the GC to $J_{\rm eff}$ scales as ${\Delta s} \times \exp(-{r}/{l_{\rm d}})$. At fixed observing angle $\beta$, this term is maximal for an aperture $\theta_{0}$ such that $\sin\theta_{0} = (r_{\odot}/l_{\rm d}) \sin \beta$. This condition requires the decay length to be larger than $r_{\odot} \sin \beta$. For smaller values of $l_{\rm d}$, the GC has little influence and the bump vanishes.
If we apply our reasoning to the dashed purple curve in the left panel of Fig.~\ref{Jfctr:smear_m_phi}, we expect the bump to be maximal for a mediator mass of $140$\,GeV while the actual value is $70$\,GeV, suggesting that other effects are at play. At fixed $\beta$, there is actually a competition between mediator decay, which favors large values of $\theta_{0}$, and the fact that too wide cones mostly shoot in empty space even if they cross the GC. The latter effect takes over the former, with the consequence that a bump appears for a smaller aperture with $\sin\theta_{0} \sim (r_{\odot} \sin \beta)/(2 \, l_{\rm d})$.
The same argument can be applied to explain the decrease of the curves beyond the bumps for large values of $m_{\phi}$. As the mediator mass increases, so does the aperture $\theta_{0}$ and the cones widen, probing regions where the DM density is on average smaller than on the los. In the limit where $m_{\phi} \simeq m_{\chi}$, we reach the non-relativistic regime where the $J$-factor is well approximated by equation~(\ref{eq:J_factor_NR_mediator}).

\begin{figure}[t!]
\centering
\includegraphics[width=0.488\textwidth]{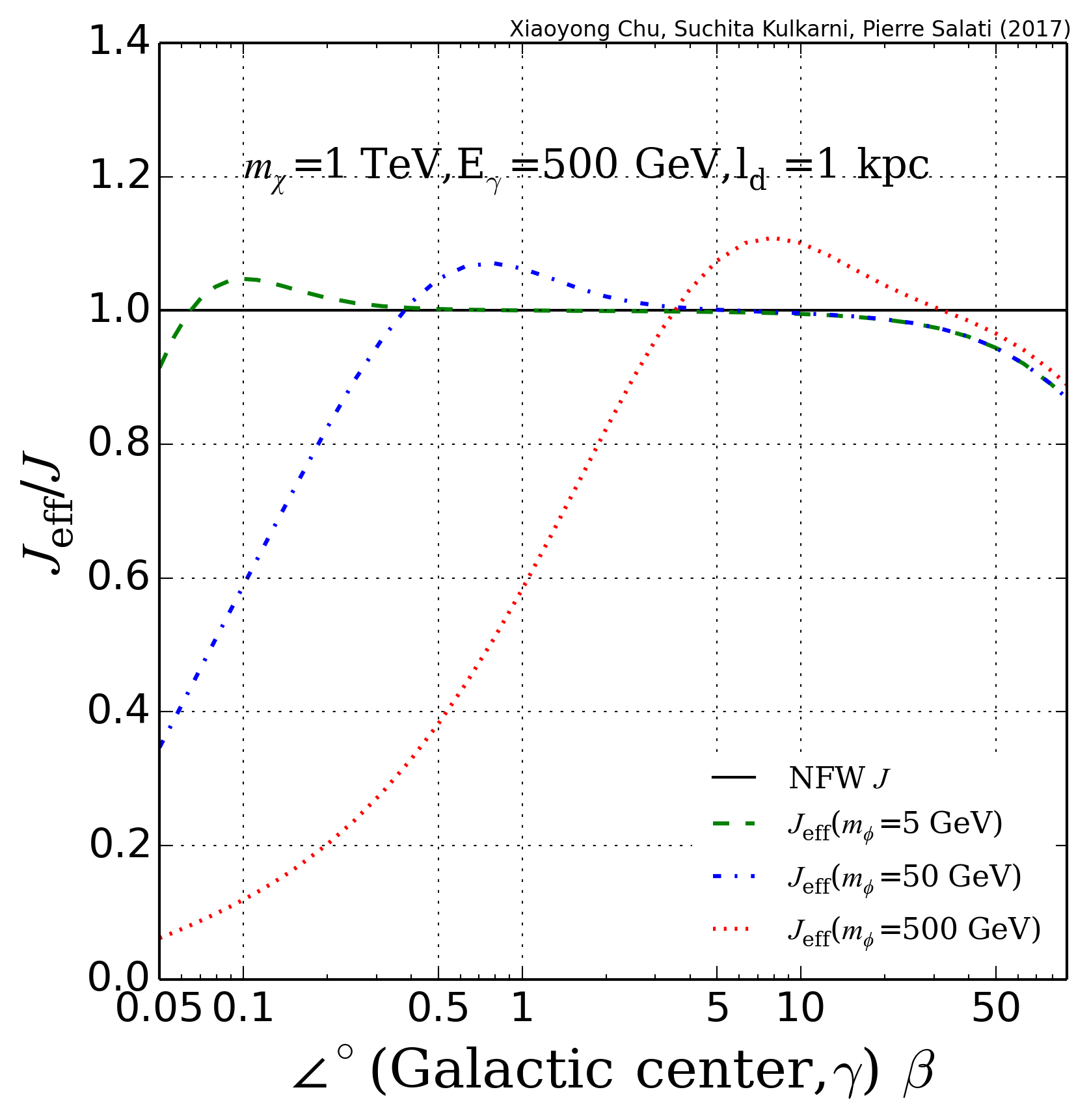}
\includegraphics[width=0.50\textwidth]{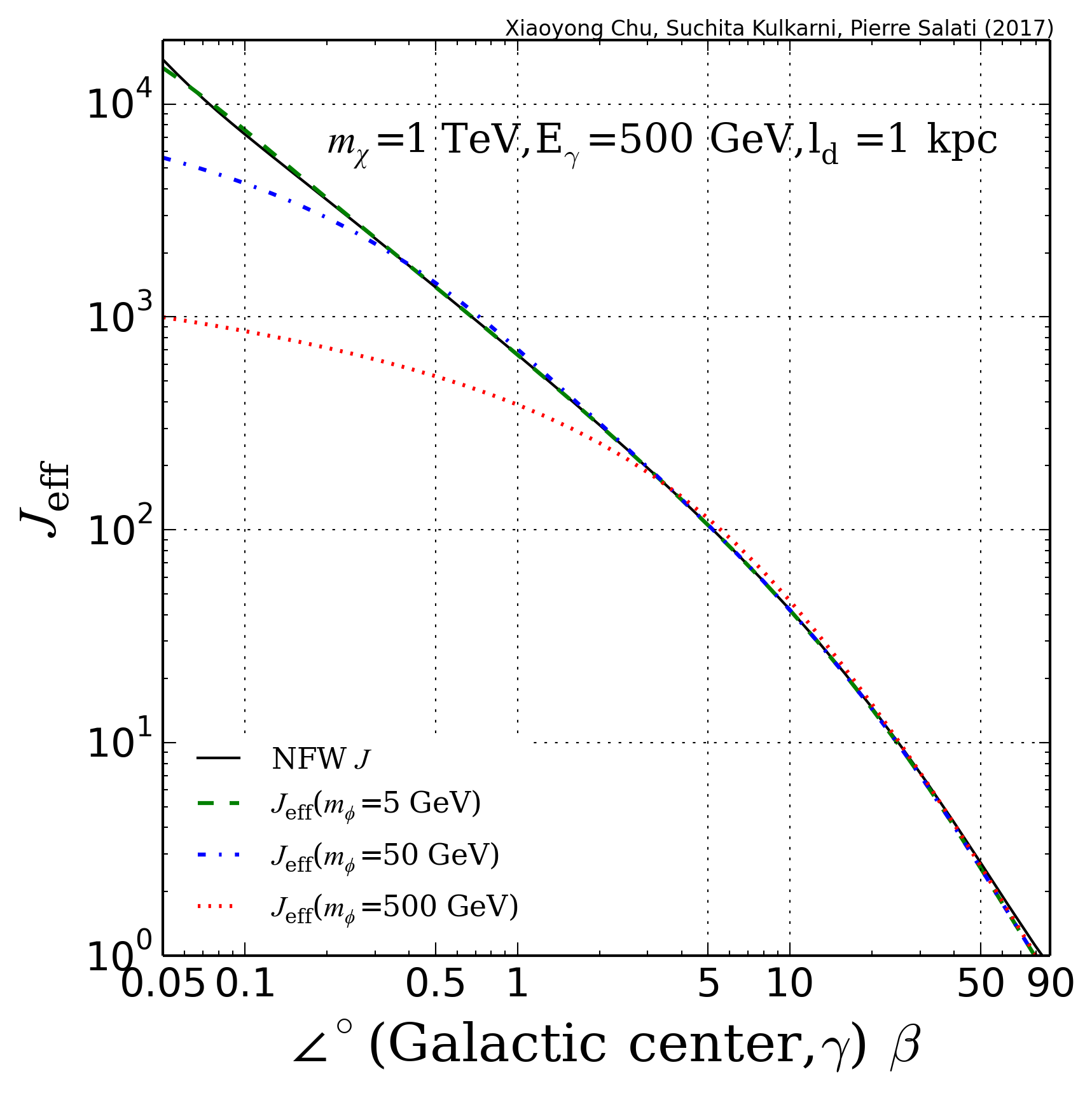}
\caption{
Dependence of the effective $J$-factor on observing angle $\beta$ for several values of mediator mass. The decay length $l_{\rm d}$ has been set equal to 1\,kpc. The left panel features the ratio of $J_{\rm eff}$ to the conventional NFW expectation $J$. In the right panel, the dimensionless value of $J_{\rm eff}$,  normalized to $r_{\odot} \, \rho^{2}_{\chi}(\odot)$, is displayed, with clear evidence for a flattening of the angular profile as the mediator mass increases. The DM mass is taken to be 1\,TeV and the observed gamma ray energy is $500$\,GeV.}
\label{Jfctr:smear_beta_1}
\end{figure}

\begin{figure}[t!]
\centering
\includegraphics[width=0.488\textwidth]{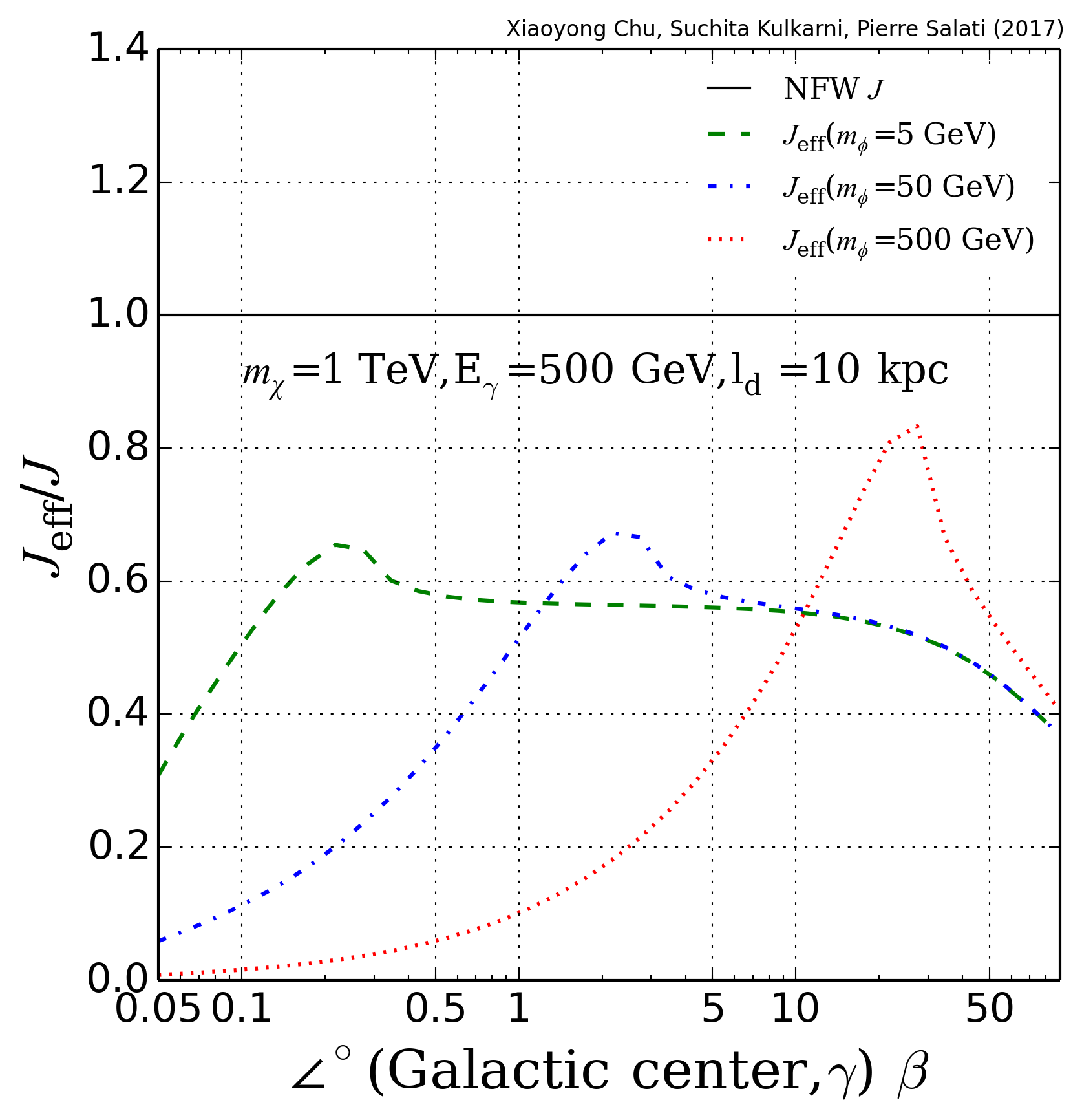}
\includegraphics[width=0.50\textwidth]{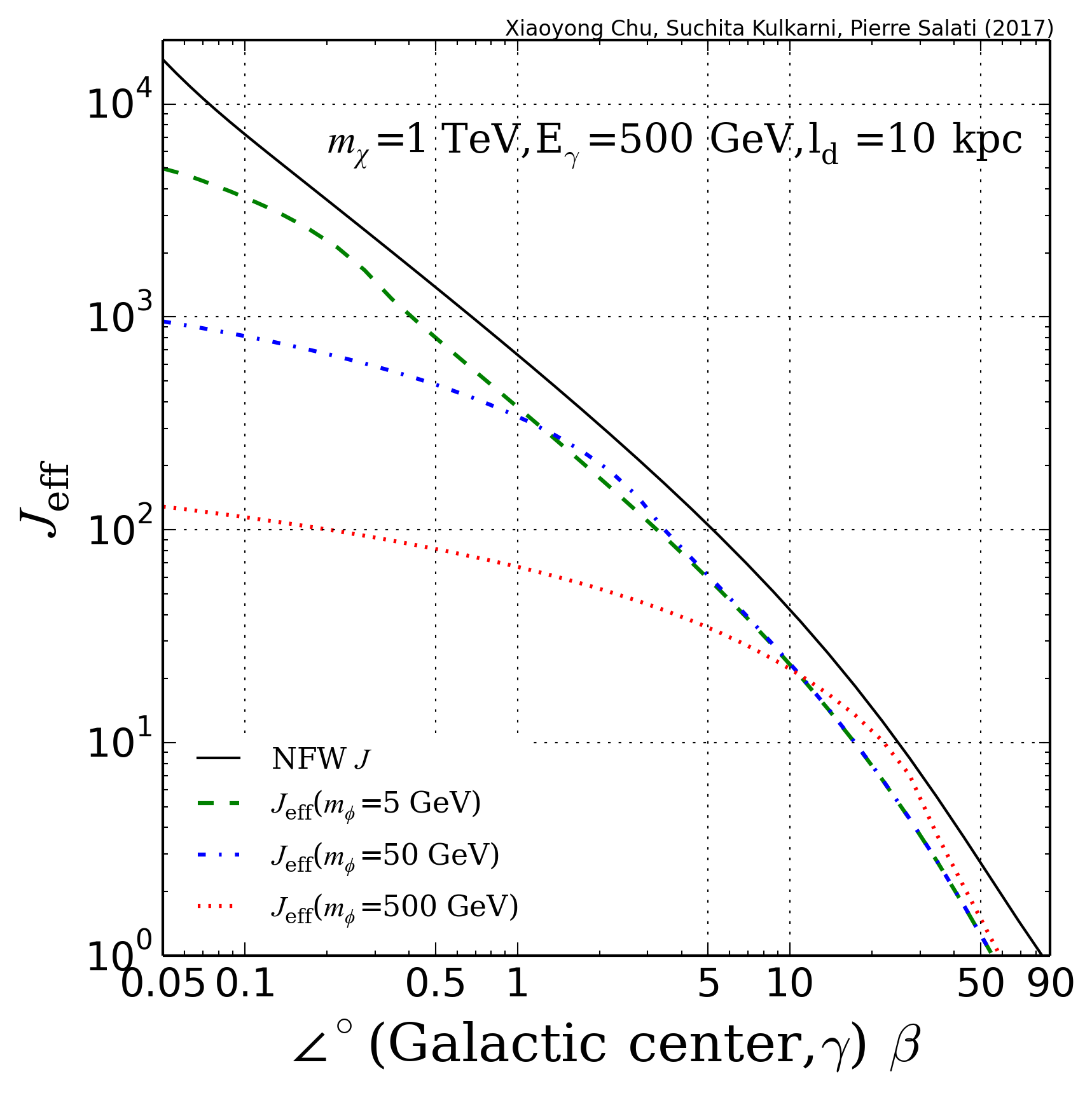}
\caption{
Same as in Fig.~\ref{Jfctr:smear_beta_1} with a decay length of 10\,kpc.}
\label{Jfctr:smear_beta_2}
\end{figure}

In Fig.~\ref{Jfctr:smear_beta_1} and \ref{Jfctr:smear_beta_2}, the absolute value of $J_{\rm eff}$ normalized to a constant $r_{\odot} \, \rho^{2}_{\chi}(\odot)$ (right panels) and its ratio to the canonical expectation $J$ (left panels) are displayed as a function of the observing angle $\beta$, for three values of the mediator mass. The decay length is respectively equal to $1$ and $10$\,kpc. The curves can be understood with the same arguments as above.
In the left panels, they exhibit a plateau which corresponds to the approximation~(\ref{eq:J_factor_UR_mediator}) for ultra-relativistic mediators. The $J_{\rm eff}$ to $J$ ratio is close to 1 (0.57) for $l_{\rm d} = 1$ (10) kpc.
Bumps are also clearly visible. Equipped with the notions which we have just discussed, we anticipate their positions to be given by $\beta_{\rm \, bump} \sim \arcsin ({m_{\phi}}/{m_{\chi}})$ in Fig.~\ref{Jfctr:smear_beta_2} where the decay length is larger than the galactocentric distance. This leads to the values of $0.28^{\circ}$, $2.8^{\circ}$ and $30^{\circ}$ in good agreement with the numerical results. As already mentioned, the actual angle is slightly smaller than our estimate.
In Fig.~\ref{Jfctr:smear_beta_1}, the decay length is smaller than $r_{\odot}$ and the damping factor from mediator decay comes into play. The positions of the bumps are now such that
\be
\beta_{\rm \, bump} \, \simeq \, \arcsin \left\{
\left( {\displaystyle \frac{2 \, l_{\rm d}}{r_{\odot}}} \right)
\left( {\displaystyle \frac{m_{\phi}}{m_{\chi}}} \right) \right\} \;,
\ee
yielding the values of $0.07^{\circ}$, $0.7^{\circ}$ and $7^{\circ}$ for a mediator mass respectively equal to $5$, $50$ and $500$\,GeV. The agreement is excellent for the blue and red curves. When $m_{\phi} = 5$\,GeV, the maximum of the ${J_{\rm eff}}/{J}$ ratio is reached for $\beta = 0.09^{\circ}$ instead of $0.07^{\circ}$.

The last and foremost property appears clearly in the right panels of Fig.~\ref{Jfctr:smear_beta_1} and \ref{Jfctr:smear_beta_2}, where the $J$-factor alone is plotted as a function of observing angle. For both values of the decay length, the angular profiles flatten out as the mediator mass increases from $5$ to $500$\,GeV.
This can be easily understood by the ``shooting across vacuum'' argument. Close to the GC, we expect the conventional NFW $J$-factor to diverge as ${1}/{\beta}$, a behaviour which is actually exhibited by the solid black lines. With mediated DM, the cones over which $J_{\rm eff}$ is calculated completely miss the GC when their aperture $\theta_{0}$ is larger than the observing angle. This occurs for values of $\beta$ smaller than $\beta_{\rm \, bump}$ as confirmed in the left panels by the drop of ${J_{\rm eff}}/{J}$ when $\beta$ goes to 0.
The flattening of the $J$-factor profile has considerable consequences. For non-relativistic mediators, a DM concentration is no longer associated to a hot spot in the gamma ray sky. By smearing their DM distributions, mediators can suppress the signal from the GC and dwarf spheroidal galaxies, lessening the bounds set by gamma ray observations on DM properties, and alleviating the possible tension arising from these constraints.

\section{Cosmic-ray positron anisotropy with mediators}
\label{sec:anisotropy}

In this section, we turn to the anisotropy induced by the long-lived mediator in cosmic-rays. For exemplification, we focus on the case that the mediator decays to a pair of electron-positron, leading to cosmic-positron anisotropy. Such anisotropy appears even if the mediator decays isotropically in its rest frame, as long as the decay  products  receive a Lorentz boost from the momentum of the mediator in Galactic frame, and thus become ``focused" along the same direction. 
Meanwhile, one should keep in mind that energetic charged particles get diffused through random scatterings on the turbulent Galactic magnetic field. Therefore, only positrons produced from mediator decays in the vicinity of the Solar System enhance the observable cosmic-positron anisotropy, while those produced outside undergo many scatterings and their momenta become isotropic rapidly. We denote such a local volume as the ``last scattering'' or ``ballistic'' sphere, characterized by the mean free path of high-energy positrons, or equivalently, the typical diffusion length, $r_{\rm max}$. In general,  $r_{\rm max}$ is a function of the final observed kinetic energy, determined by the propagation model of cosmic rays. 

Below we start with necessary definitions for cosmic positron anisotropy.  Then we apply them to both two-body and three-body decay cases, and discuss the prospects for their experimental detection. Our results can also be readily applied to photons, which propagate freely in space, by setting the diffusion length $r_{\rm max}$ to be infinite.

\subsection{Formalism for positron anisotropy}
\label{subsec:form_pos_ani}

As stated above, to calculate the positron anisotropy, one needs to know the positron flux produced by mediator decay both inside and outside the ballistic sphere. On the one hand, the positron flux produced within the ballistic sphere, i.e. $\Phi_{e}(\vec{w}, E_{e})$, can be derived directly from the general expression equation~\eqref{eq:gam_flux_3} by replacing photons with positrons.
On the other hand, positrons produced outside this sphere are supposed to be sufficiently diffused and thus are treated as the homogeneous background here, denoted as $\Phi^{\rm diff}_{e}(E_{e})$. This quantity is obtained by applying the effective density $\rho_{\rm eff}$ of equation~\eqref{eq:rho_eff} to the propagation model of cosmic rays. At last, we have assumed that the astrophysical background of secondary positrons is only sub-leading and will not be considered here. A more realistic treatment is left for future work. 

Then, given a certain observed positron energy $ E_{e}$, one can define the angle-dependent anisotropy as follows:
\begin{equation}
\Delta(\vec{w}, E_{e})\, \equiv \, {\Phi_{e}(\vec{w}, E_{e}) \over  \Phi^{\rm total}_{e}(\vec{w}, E_{e})} \,\sim\, {\Phi_{e}(\vec{w}, E_{e}) \over  \Phi_{e}(\vec{w}, E_{e}) + \Phi^{\rm diff}_{e}(E_{e})},
\end{equation}
where  we have adopted the  so-called ``MAX" model for the propagation of cosmic positrons in the Galaxy. More concretely, the numerical parameters of the propagation model are taken to be ${\delta}$=0.46, $\tau_E$=10$^{16}$\,sec, $K_0$=0.0765\,kpc$^2/$Myr, the diffusion coefficient K(E)= K$_0(v/c)(E/\text{GeV})^{\delta}$, and the energy loss coefficient $b(E)$= E$^2$/(GeV$\cdot \tau_E$)~\cite{Donato:2003xg}. In the end, the typical diffusion length of relativistic positrons can be estimated by 
\begin{equation}
	r_{\rm max}(E_e) \simeq \frac{ 3 K(E_e)}{c} \simeq 0.73\text{\,pc}\times  \left({E_e\over \text{GeV}}\right)^{\!\delta}.
\end{equation}  
For the observed positron energy  of $ 50$\,GeV and $500$\,GeV, it gives  4.4\,pc and 12.7\,pc, respectively. 

To directly compare our results with experimental limits on dipole anisotropies, we further expand the fluctuation  $$ { \Phi^{\rm total}_{e}(\vec{w}, E_{e}) - \langle \Phi^{\rm total}_{e}\rangle \over  \langle \Phi^{\rm total}_{e}\rangle } = \sum_{lm} a_{lm}(E_{e}) Y^l_m(\vec{w}), $$ in the basis of spherical  harmonics $Y^l_m(\hat n)$. Following  \cite{Abdollahi:2017kyf},  the dipole anisotropy of positron flux can then be estimated by 
\begin{equation}
{\mathcal A}(E_e) \equiv {3 \over  \sqrt{4\pi}} \times \sqrt{ \sum_{m} |a_{1m}^2 | \over 3 } = 3\times {\left| \int^1_{-1} d\cos\beta \, \Phi^{\rm total}_{e}(\vec{w}, E_{e})\cos\beta \right|\over \int^1_{-1} d\cos\beta \, \Phi^{\rm total}_{e}(\vec{w}, E_{e})}  , 	\label{dipole:aniso}
\end{equation}
where $\beta$ is the observing angle defined in the previous section, i.e., the angle between the los and the GC.

Before diving into detailed calculations of the positron anisotropy for concrete models in the following subsections, we would like to recall the status of relevant experimental observations. Seven years of Fermi-LAT data constrain dipole anisotropy of $(e^+ + e^-)$ to be below ${\mathcal O}\,(10^{-2})$ for positron energy, $E_e $, at the electro-weak scale~\cite{Abdollahi:2017kyf}. Provided that cosmic positrons make up more than 10\% of  $(e^+ + e^-)$  spectrum above $100$\,GeV, this can be interpreted  as an upper bound on positron anisotropy at the order of ${\mathcal O}(10^{-1}) $ if spatial fluctuations in electron flux do not accidentally compensate positron anisotropy.  At the same time, PAMELA~\cite{Adriani:2015kfa}, and AMS-02~\cite{LaVacca:2016tqq} collaborations have put direct constraints on cosmic positrons, suggesting upper bounds on ${\mathcal A}(E_e)$ of the range $0.02$\,-\,$0.1$ for minimal $E_e$ from 16 to 100\,GeV.  As shown below, current bounds are generally too weak to probe the scenario of a long-lived mediator, except at very high energies. 

\subsection{two-body decay}
\label{subsec:2-body}

In the case that scalar mediators decay directly into $e^{+}e^{-}$ pairs, positrons are produced in the mediator rest frame with energy $E_{e}^* = {m_{\phi}}/{2}$. Just like the previous case of photons, the final positron energy spectrum  in Galactic frame is ``box-shaped", and there also exists a one-to-one correspondence between the positron energy, $E_{e}$, and the angle between mediator and final positron momenta, $\theta$, in the Galactic frame (see equation~\ref{theta:energy}).  
It is straightforward to see that in the limit of massless positron and mediator, $\theta$ converges to zero, where the final positron moves colinearly along the direction of its parental mediator. 
The energy distribution of positrons $\Dpos(\cos\theta, E_{e}) $ is also similar to that of photons:
\be
 \Dpos(\cos\theta , E_{e})  ~=~ {\displaystyle \frac{{m_{\phi}^{2}}/{m_{\chi}^{2}}}{4 \pi (1 - \beta_\phi \cos\theta)^{2}}} \times
\delta \! \left\{E_{e} - {\displaystyle \frac{{m_{\phi}^{2}}/{2 m_{\chi}}}{(1 - \beta_\phi \cos\theta)}} \right\} \,, \label{distr:twobody}
\ee
for which a more detailed derivation can be found in appendix~\ref{app:positron}. 

The anisotropic flux depends on the phase space distribution $\psdphi(\vec{x}_\odot , \vec{u})$, and in general is dominated by mediators coming from the GC, where mediator lifetime plays an important role. In contrast, the diffuse background carries no information on mediator propagating directions, and thus directly relies on the smeared density profile $\rho_{\rm eff}$. As a result, as modifying  $l_{\rm d}$ hardly changes the local $\rho_{\rm eff}$ (see left panel of Fig.~\ref{fig:medprofile}), the diffuse flux is rather insensitive to the decay length.
We have assumed ultra-relativistic mediators and applied relation~(\ref{eq:J_factor_UR_mediator_0}) to compute the flux of the prompt positrons produced inside the ballistic sphere. The integral along the los is now performed from $0$ to $r_{\rm max}$. Two-body decay results into a ``box-shaped" energy spectrum and leads to the flux
\be
\Phi_{e}(\vec{w}, E_{e}) \, \simeq \,  {\displaystyle \frac{1}{4 \, \pi}} \,
{\displaystyle \frac{\sv}{m_{\chi}^{2}}} \, {\displaystyle \frac{dN_{e}}{dE_{e}}} \times {\displaystyle \frac{r_{\rm max}}{l_{\rm d}}} \times
\left\{ J_{\phi}(- \vec{w}) \equiv {\displaystyle \int_{0}^{+\infty}} dr \; \rho_{\chi}^{2}(\vec{x}_{S}) \, e^{\displaystyle {-r}/{l_{\rm d}}}
\right\},
\ee
which scales like
\be
\Phi_{e}(\vec{w}, E_{e}) \propto {E_{e}^{\,\delta} \over \sin\beta} \times {1 \over l_{\rm d}} \, e^{-{r_{\odot}/{l_{\rm d}}}} \;.
\label{nondiffuse:flux}
\ee
The diffuse positron flux is calculated according to the usual pseudo-time procedure~\cite{Baltz:1998xv} and may be expressed as the convolution over energy and magnetic halo (MH) of the positron propagator ${\cal G}_{e^{\! +}}$ and positron source term $q_{e^{\! +}}$
\be
\Phi^{\rm diff}_{e}(E_{e}) \, = \, {\displaystyle \frac{c}{4 \, \pi}} \,
{\displaystyle \int_{E_{e}}^{m_{\chi}}} dE \,
{\displaystyle \int_{\rm MH}} d^{3}\vec{x} \times
{\cal G}_{e^{\! +}} \! \left\{ \vec{x} , E \rightarrow \odot , E_{e} \right\} \times q_{e^{\! +}}(\vec{x} , E) \;. \label{diffuse:flux_exact}
\ee
The injection rate of positrons by mediator decays at point $\vec{x}$ has been calculated in section~\ref{sec:eff_DM_rho} with the help of relation~(\ref{eq:ann_DM_eff}) where the smeared density $\rho_{\rm eff}$ is given by definition~(\ref{eq:rho_eff}). An observer collects at the Earth the positrons produced only inside the so-called ``positron sphere'' which, at the energies of interest, extends at most over a few kpc. We may calculate the positron propagator as if the MH were infinite. Moreover, the effective DM density $\rho_{\rm eff}$ is approximately constant over the ``positron sphere'', allowing us to simplify the previous expression into
\be
\Phi^{\rm diff}_{e}(E_{e}) \, \simeq \, {\displaystyle \frac{1}{4 \, \pi}} \,
{\displaystyle \frac{\sv}{m_{\chi}^{2}}} \, {\displaystyle \frac{dN_{e}}{dE_{e}}} \times {\displaystyle \frac{c}{b(E_{e})}} \times 
\left( m_{\chi} - E_{e} \right) \times \rho_{\rm eff}^{2}(\odot) \propto {{m_{\chi} - E_{e}}\over{E_{e}^{2}}} \; . \label{diffuse:flux}
\ee
The behaviour of positron anisotropy can be understood as
\begin{equation}
\Delta(\vec{w}, E_{e})\, \propto {E_e^{\,2+\delta} \over  \sin\beta \left( m_{\chi} - E_{e} \right)} \times  {1\over l_{\rm d} }\, e^{-{r_{\odot}/{l_{\rm d}}}}.\label{Delta2:approx}
\end{equation}

\begin{figure}[t!]
\centering
\includegraphics[width=0.50\textwidth]{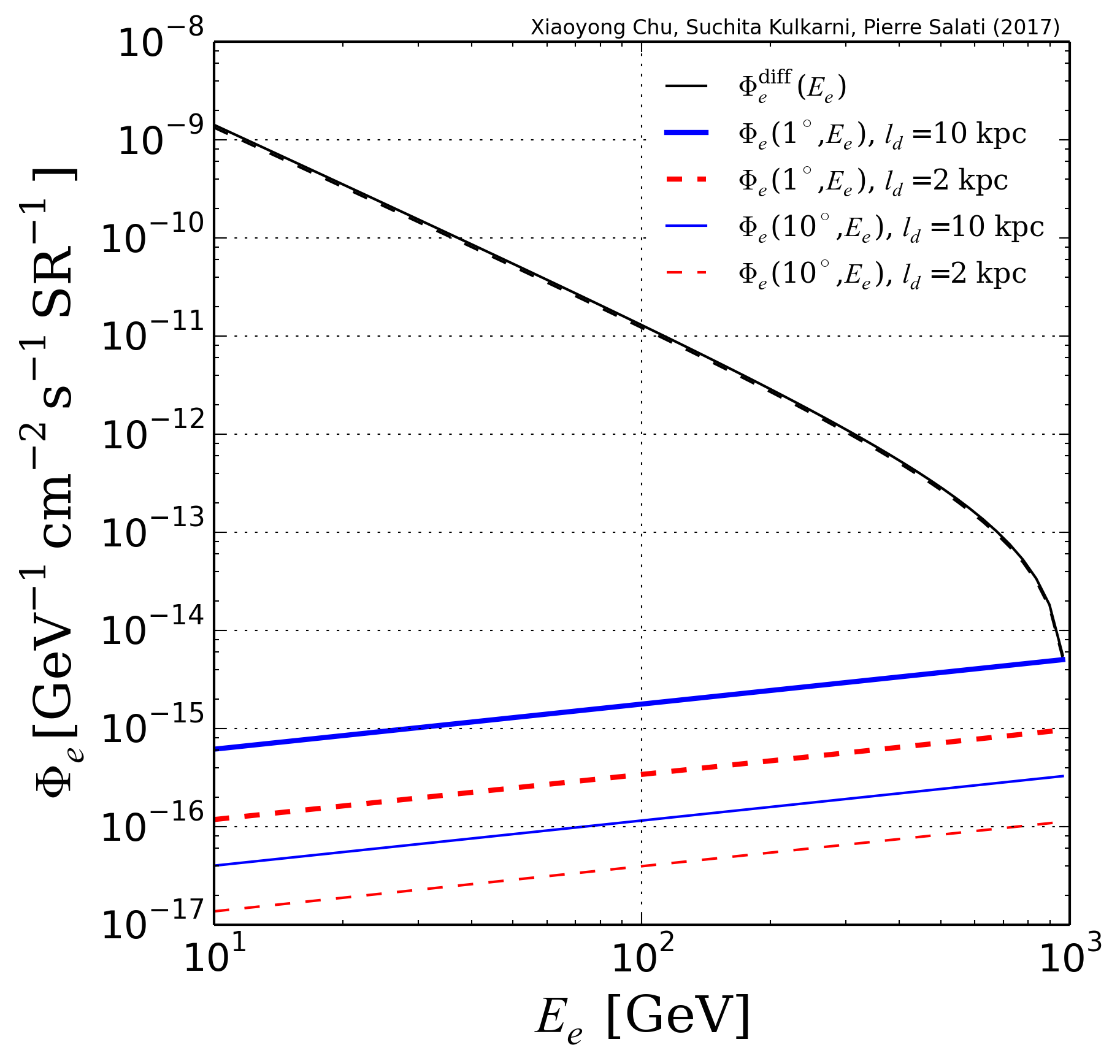}\includegraphics[width=0.485\textwidth]{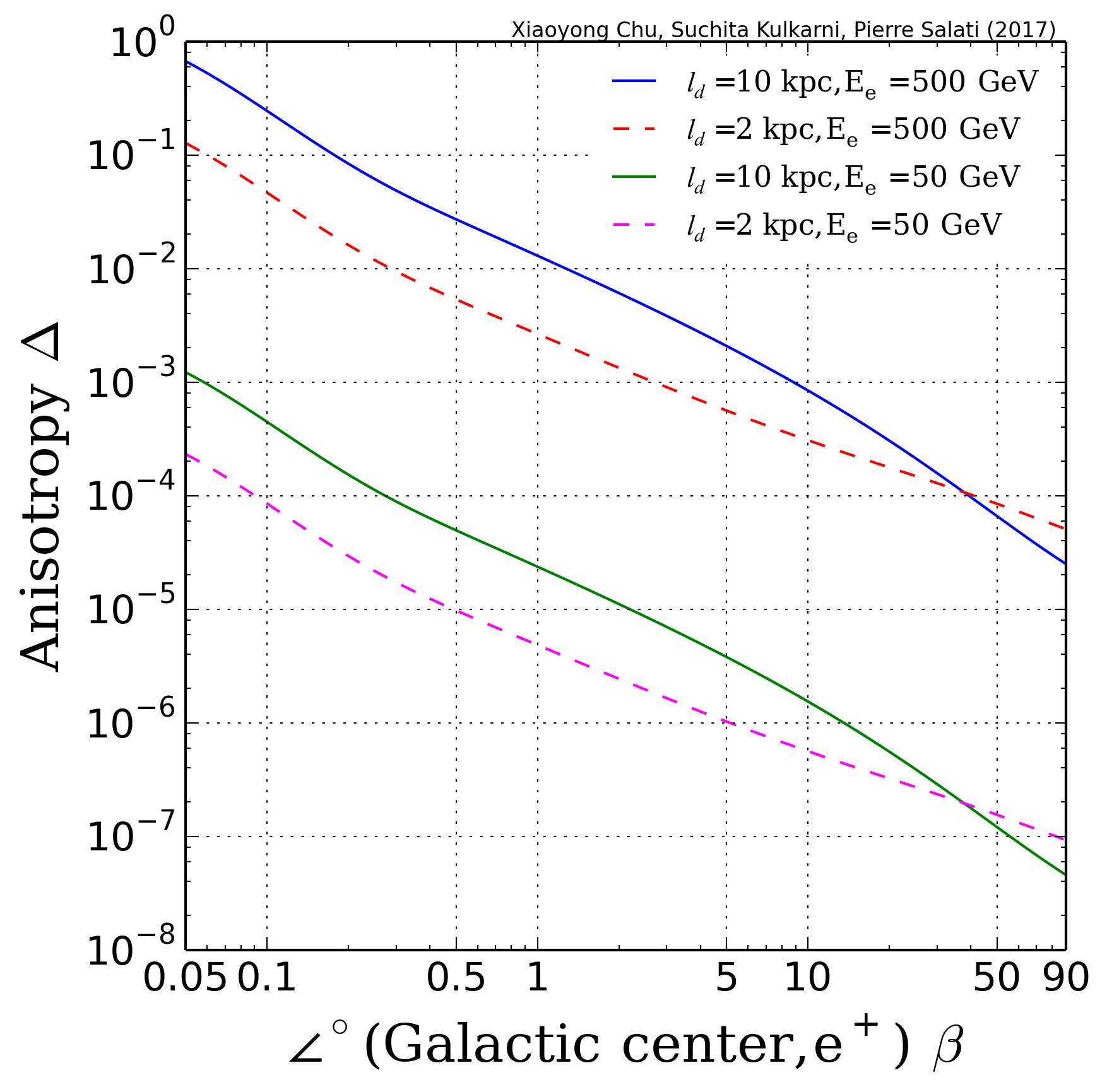}
\caption{
Left panel: Computed fluxes of cosmic positrons as a function of observed positron energy. Black curves show the diffuse component while colored lines feature the prompt flux.
Right panel: Positron anisotropy  $\Delta(\vec{w}, E_{e})$  as a function of observing angle $\beta$.  Both panels are for 1\,TeV WIMP with the NFW profile as given above, and solid  (dashed) curves correspond to a decay length $l_{\rm d}$ of $10$\,kpc ($2$\,kpc). Here the mediator $\phi$ decays to a pair of electrons, with a ``box-shaped'' energy spectrum.}
\label{DMDA:NFWprofile}
\end{figure}

To verify these analytical expressions, we calculate numerically the anisotropies for the NFW DM profile, whose parameters have been introduced earlier, and show the computed results in Fig.~\ref{DMDA:NFWprofile}.
This is done for 1\,TeV DM, where we take two benchmark decay lengths: 10\,kpc   and 2\,kpc, illustrated by solid and dashed curves, respectively, while varying the observed positron energy (left panel) and the observing angle (right panel).

In Fig.~\ref{DMDA:NFWprofile}, the left panel gives the diffuse (prompt) flux of cosmic positrons as black (colored) curves, respectively, as functions of the observed positron energy. The diffuse flux simply decreases with $E_{e}$ as described in equation~\eqref{diffuse:flux}, with little dependence on $l_{\rm d}$. The situation is different for the prompt flux, where increasing $E_{e}$ raises the diffusion length $r_{\rm max}$, and thus the value of $\Phi_{e}(\vec{w}, E_{e})$. It in turn enhances the dipole anisotropy signal, $\Delta$.  This effect is shown in the right panel, where the anisotropy values of the upper set of two curves, corresponding  to $E_e=500$\,GeV, is about a factor of $500$ larger than those of the lower two with $E_e=50$\,GeV, in agreement with the  approximation given by equation~\eqref{Delta2:approx}.  It is worth mentioning that the increase of positron anisotropy with the energy threshold $E_e$ comes with the price of a significantly reduced number of total detectable events.

Because  the prompt positron flux is usually dominated by the contribution of mediators produced in GC, larger $\Phi_{e}(\vec{w}, E_{e})$ can also be achieved by either increasing  $l_{\rm d}$ or reducing $\beta$, as can be seen from right panel of Fig.~\ref{DMDA:NFWprofile}. These effects are rather mild except for very small observing angle,  and to observe the latter in observatories would require high angular resolutions. For large observing angles, it  becomes less likely to observe events induced by mediators from the GC, so the total signal reduces significantly. Another consequence is that the dependence of $\Delta$ on $l_{\rm d}$ becomes much weaker and each set of curves tends to converge with increasing observing angle.    

With respect to the dipole anisotropy, for  1\,TeV  DM we obtain  ${\mathcal A}(E_e= 50\text{\,GeV}) \simeq 2.37\times 10^{-7}$ and ${\mathcal A}(E_e= 500\text{\,GeV})\simeq 1.30\times 10^{-4}$,  respectively, from equation~\eqref{dipole:aniso}. The decay lifetime has been chosen to yield  $l_{\rm d}= r_{\odot}$ to maximize the results.  These small numbers are well beyond the experimental sensitivity at this moment. Therefore, it is phenomenologically more interesting to look for small-scale anisotropies, especially  $\Delta(\vec{w}, E_{e})$ with very small observing angle.

At last, we briefly comment on how  the observable signal can be enhanced from the aspect of particle physics. One possibility is to consider  small-scale anisotropies in gamma rays, which travel freely in space, i.e. $r_{\rm max}$ goes to infinity, as studied in the last section.  Although the mediator dominantly decays to electrons, gamma-ray anisotropies can be induced by both final state radiation of the mediator decay and inverse Compton scattering of mediator positrons with interstellar radiations~\cite{Ibarra:2009nw}.  Another possibility is to introduce  mediators with non-zero spin, which changes the energy spectrum of final SM products\,\cite{Garcia-Cely:2016pse}, and thus may be used to increase the observed anisotropy at higher energies without losing too many observable events.

\subsection{three-body decay}
\label{subsec:3-body}

In this subsection, we turn to another possibility that the mediator decays into a pair of electron-positron plus another lighter dark particle, following the proposal of \cite{Kim:2017qaw}. The mass ratio of the final light dark state to the mediator, $R$, is regarded as a free parameter here.  Both the mediator and the final dark particle are set to be scalars, so the mediator decays isotropically. We further assume positron to be massless for simplicity. Given the energy spectrum of positrons in the rest frame of the mediator~\cite{Fortin:2009rq}, we obtain the  distribution of positron energy as
\be
 {\mathcal D}_e (\cos \theta, E_e)  = {1\over 4\pi \Gamma_\phi(1-\beta_\phi \cos\theta) }\times  \left\{ {dN_e\over dE^*_{\rm e}} \equiv   {8 E^*_{\rm e} (m_\phi -R^2 m_\phi -2 E^*_{\rm e}) \over m_\phi^2 (m_\phi -2 E^*_{\rm e})(1+4R^2\log[R]-R^4)} \right\},\label{three:spectrum}
 \ee
where $E_e^* \equiv \Gamma_\phi (1-\beta_\phi \cos\theta)E_e $. More details of $(\cos\theta , E_{e})$ transformation rules  between two frames are given in appendix~\ref{app:positron}.

Energy-momentum conservation requires that the maximal positron energy in mediator-rest frame, $E_e^*$, is $(1-R^2)m_\phi/2$, suggesting that kinetically allowed values of $\theta$ should be relatively small. In practice,  the large mass hierarchy suggests the collinearity between the momentum directions of final electrons and high-energy $\phi$, i.e. $\theta \simeq 0$. Nevertheless, we have solved the exact values of $\theta$ for the numerical results below.

To test this scenario as a potential DM explanation of PAMELA/AMS positron excess,  we assume a very dense core in the GC, and the usual NFW profile outside, of 1\,TeV DM particles~\cite{Kim:2017qaw}. The formation of such a core can be caused by black hole adiabatic formation, as has been discussed in section~\ref{sec:blackhole}. While the diffuse flux is insensitive to the size of the core, its radius changes the morphology of small-scale anisotropy dramatically. Here we consider both very small core ($1$\,pc and ${\mathcal N}\equiv \rho_{\rm c}/\rho_{\rm NFW,\,c} = 5900$) and very large core ($0.5$\,kpc and ${\mathcal N}=277$), while black hole contraction suggests some value between, about tens of pc from section~\ref{sec:blackhole}. The property of the mediator is as described above. Similar to \cite{Kim:2017qaw}, we set $R=0.2$ and the decay length of the mediator at two benchmark values:  2\,kpc and 10\,kpc. 

\begin{figure}[t!]
\centering
\includegraphics[width=0.47\textwidth]{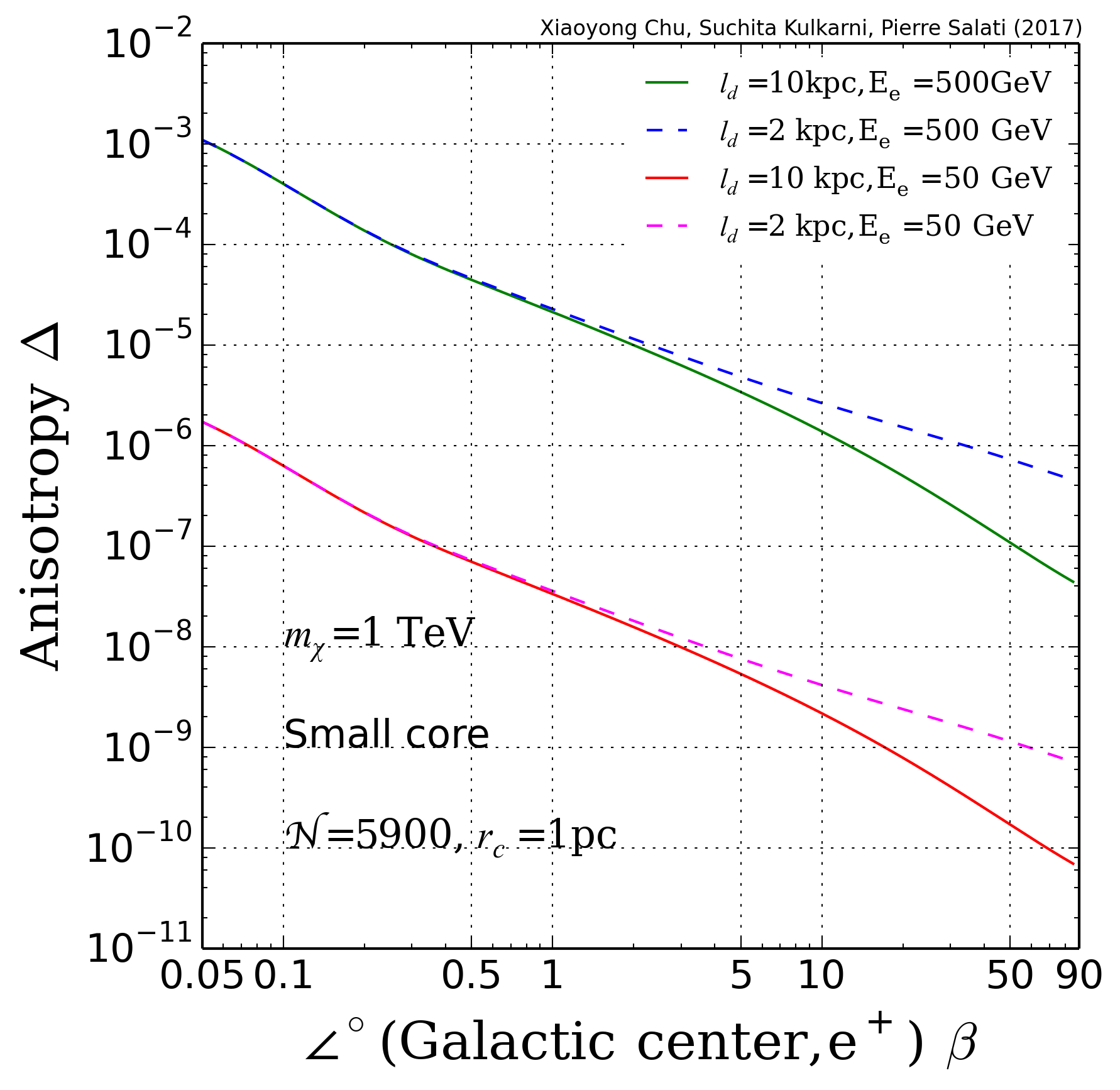}
\includegraphics[width=0.47\textwidth]{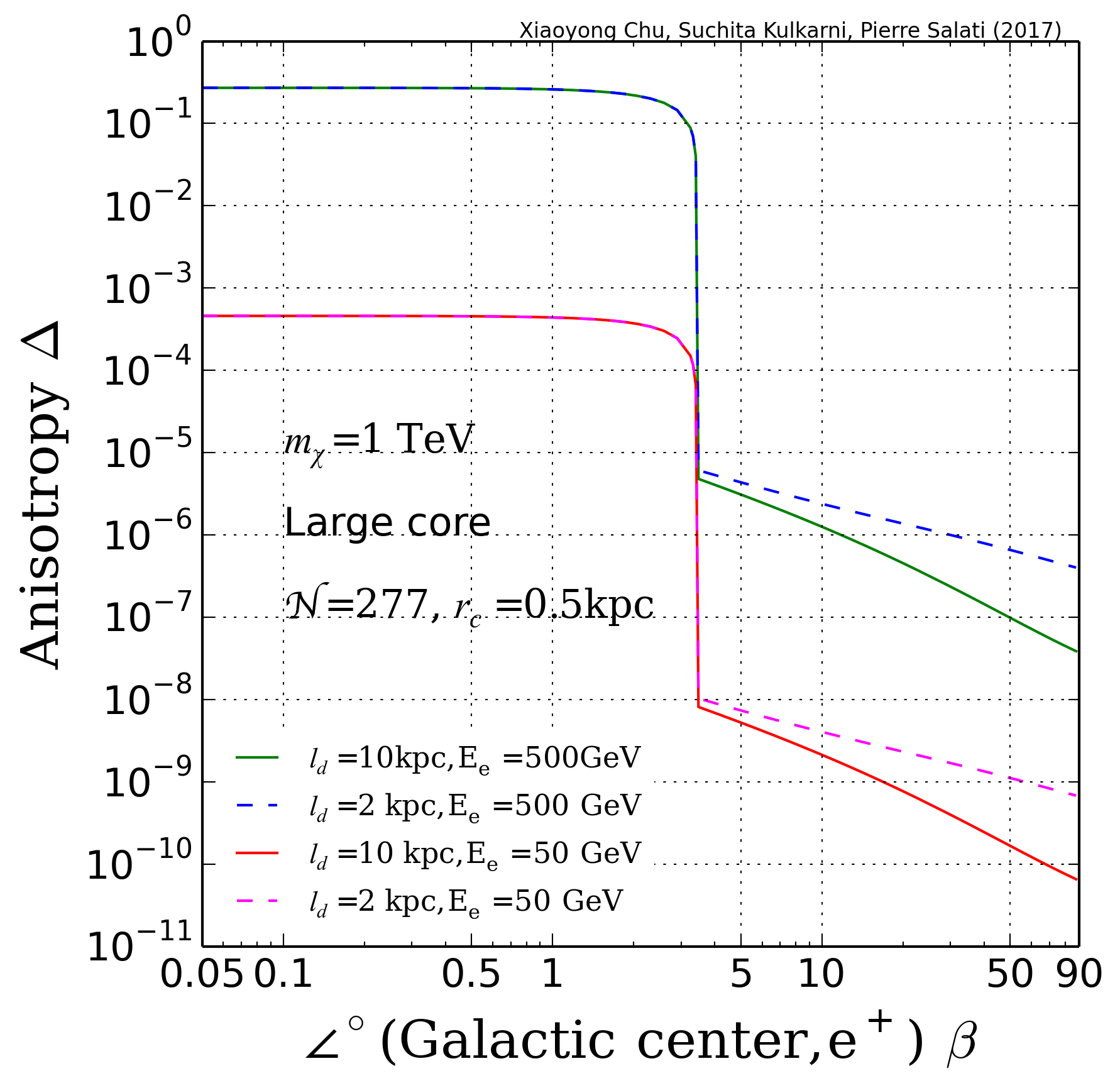}
\caption{Positron anisotropy  $\Delta(\vec{w}, E_{e})$  as a function of observing angle $\beta $ for both a small core (left panel) and large core (right panel). See texts for more details of the DM profile. Here the mediator $\phi$ decays to a pair of electron plus one light dark state, and energy spectrum is given by equation~\eqref{three:spectrum}.  }
\label{DMDA:Densecores}
\end{figure}

Fig.~\ref{DMDA:Densecores} shows the anisotropies with those input parameters above, as functions of observing angle $\beta$.   On the one hand, as demonstrated in the last subsection and implied by equation \eqref{nondiffuse:flux}, the prompt flux of positrons is approximately dominated by DM annihilation events taking place in the inner region of dark halo. On the other hand, the extremely dense core we considered in this subsection, together with $l_{\rm d}$ of the order of several kpc, also decides the diffuse positron flux\footnote{Approximating equation \eqref{diffuse:flux_exact} for infinite magnetic halo, as done in this work, we get one order of magnitude difference for the diffused flux with respect to \cite{Kim:2017qaw}. On the one hand, this discrepancy does not impact the cosmic ray anisotropy. On the other hand, detailed understanding of the situation will require numerical simulation, which is beyond the scope of current work and will be commented upon in the future.}. This has already been demonstrated by equation \eqref{enhance:diffuse} and shown in the right panel of Fig.~\ref{fig:medprofile}. As a result, the dependences of both fluxes on the decay length cancel out in computing the anisotropy, as long as both approximations above work well. This observation explains the similarity of solid ($l_d =10\,$kpc) and dashed ($l_d =2\,$kpc) curves in Fig.~\ref{DMDA:Densecores}, if other parameters are the same. 
Even though, large prompt and diffuse fluxes are preferred in order to explain the positron excess, and for the purpose of observation. The maximal fluxes are reached at $l_d = r_\odot$, with its half-maximum range between 3.1 and 35.4\,kpc.
Another important feature in the figure is that in the case of relatively large observing angles, $i.e.$ $\beta > 0.007^\circ$ for $r_c = 1\,$pc (left panel, outside the plotting range) and  $\beta > 3.5^\circ$ for $r_c =0.5\,$kpc (right panel), DM annihilation  inside the dense core can not contribute to the prompt flux any more, so the anisotropy decreases dramatically, as illustrated in the right-bottom corner of right panel in Fig.~\ref{DMDA:Densecores}.
This is different from positrons created by nearby pulsars. In the latter case, the small-scale anisotropy only changes very mildly with the observing angle due to the fact that there is no prompt positron flux induced by long-lived mediators~\cite{Manconi:2016byt}.

\begin{table}[h]
{\centering 
\begin{tabular}{| l | c | c |} 
\specialrule{.5pt}{0pt}{0pt}
 1\,TeV DM & $E_{e}$ = 50\,GeV  & $E_{e} $= 500\,GeV    \\ \specialrule{.5pt}{0pt}{0pt}
 small core (1\,pc) &  $8.45\times 10^{-7}$ &   $5.37\times 10^{-4}$ \\
large core (0.5\,kpc) &   $8.44\times 10^{-7}$ &   $5.36\times 10^{-4}$ \\
\specialrule{.5pt}{0pt}{0pt}
\end{tabular}
}
\caption{Dipole anisotropies for mediator decay length $l_d=10$\,kpc at two different energies for DM mass of 1\,TeV. }
\label{table:ani}
\end{table}

At last, we also calculate the dipole anisotropy, ${\mathcal A}(E_e) $, for $l_{\rm d} = 10$\,kpc, which is given in  table~\ref{table:ani}. 
Such level of dipole anisotropies may become detectable in the near future, once high-energy indirect searches, such as H.E.S.S.~\cite{Abdalla:2016olq} and CTA~\cite{Carr:2015hta} experiments,  are able to collect  enough event samples\footnote{The situation is similar to that of using local sources to explain positron excess. See \cite{Abdollahi:2017kyf} and references therein.}. Besides, it is easier to probe the case of a large core, from measuring both small-scale  and dipole  anisotropies.  


\section{Conclusions}
\label{sec:conclusion}
In this work, we explore the phenomenology of dark matter annihilating to long-lived mediators, which subsequently decay to SM particles in a model independent manner. We assume that the mediators are much lighter than the dark matter particles. Such scenarios have been used to explain the AMS-02 positron excess by Kim et al.~\cite{Kim:2017qaw}, following an earlier proposal by~\cite{Rothstein:2009pm}.

The first and foremost result of our analysis is that the long-lived mediator smears the dark matter density profile for indirect searches. Due to the decaying nature of the mediator, the effective density profile gets smeared by an exponential factor, attenuating the dark matter density spike in the inner regions of the galaxy and enhancing it in the outer regions. As already mentioned by~\cite{Rothstein:2009pm} in the case of the NFW distribution, such modification of  the effective profile of the annihilating dark matter mimics that of conventional decaying dark matter, regarding indirect signals of dark matter. The injection rate of SM particles is no longer directly related to the dark matter density,   allowing for a large dark matter spike at the Galactic center which could alleviate possible contradictions with the experimental constraints.  We have analyzed the possibility of acquiring a large central dark matter spike required to explain the AMS-02 excess via adiabatic black hole contraction. We argue that formation of large concentration is marginally possible under reasonable circumstances. 

In the second part of our work, we have computed the flux of prompt final state particles -- positrons and photons -- resulting from the decay of mediators. We have derived the effective $J$-factors useful for the computation of the flux, and showed that the $J$-factors in these scenarios are no longer independent of particle physics. In particular, the $J$-factor depends on the ratio of dark matter to mediator mass, i.e. the boost of the final state particles and the decay length of the mediators. Finally, after setting up  the formalism for anisotropy computation for two and three body decays of the mediator, we have computed both the small-scale and dipole anisotropies resulting from the decay of long-lived mediators. 
Current levels of anisotropy generated in such processes are small, however, they might be within the  reach of next generation indirect detection experiments. 

Dark matter annihilating to long-lived mediators thus forms an exciting avenue for indirect detection of dark matter. The model independent framework set up in this work can be readily applied to concrete particle physics models. Application of this framework to realistic models and a global analysis of multi-messenger constraints are left for future work.

\section{Acknowledgements}
\label{sec:ack}
XC and SK are supported by the `New Frontiers' program of the Austrian Academy of Sciences. XC and SK thank  Josef Pradler and Gabrijela Zaharijas for useful discussions.


\appendix

\section{Energy/Angle transformation in special relativity }
\label{sec:appendix_A}
In this section, the angle between $\vec u$ and $\vec w$  is denoted as  $\theta $ in the Galactic frame ( $\theta^*$ in the center-of-mass frame).   The positron carries energy $E_e$ in the Galactic frame (or $E_{\rm e}^*$ in the center-of-mass frame).   Note that in the Galactic frame, the positron should have a different angle, $\theta$, from the electron produced by the same decay, and carry a different energy  (except for the case that $\theta^* =\pi/2$) due to momentum/energy conservation.   To simplify the equations below, we first define two Lorentz factors: $\Gamma_\phi = m_\chi/m_\phi$ and $\Gamma_e = E_{\rm e}^*/m_e$, as well as the ratio of particle velocity to speed of light: $\beta_{\phi\,(e)}=\sqrt{1-\Gamma_{\phi\,(e)}^{-2}}$. Of course for photons $\beta_{\gamma} \equiv 1$ and $1/\Gamma_{\gamma} \to 0$.  Also note that in the case of two-body decay of mediator, $E_{\rm e}^* \equiv m_\phi/2$, while there in general exists a continuous spectrum of $E_{\rm e}^* $ for three-body decay.

Textbook knowledges of special relativity give the transformation rules  between the Galactic and center-of-mass frame of the  kinetic energy of positron:
\be
E_e = \Gamma_\phi (E_{e}^* + \vec \beta_\phi \cdot \vec P_{e}^*)  = \Gamma_\phi E_{e}^* (1 +   \beta_\phi  \beta_{e} \cos\theta^*)  \,,\label{eEnergy:CM}
\ee 
and of the angle:
\be
\tan\theta  =  {  \beta_e \,\sin\theta^*  \over  \Gamma_\phi \,( \beta_e\cos\theta^* + \beta_\phi)}\,.
\ee

Both quantities in the Galactic frame are given above as functions of the angle of $\vec\beta_\phi$ and $\vec\beta_e$ in the center-of-mass frame,  $\theta^*$. Phenomenologically, it would be more useful to have $E_e$ in terms of $\theta$. We obtain
\begin{equation}
	E_e = \Gamma_\phi  E_{e}^*\times { 1 \pm   \beta_\phi  \sqrt{|  \cos^2 \theta\,   (1 -{\cos^2\theta \over \Gamma_e^{2} } -  {    \Gamma_\phi^2 \sin^2\theta \over\Gamma_e^{2}}  )|} 
	\over 1 + \Gamma_\phi^2\beta_\phi^2 \sin^2\theta  }\, ,
\end{equation}
where ``$+$" sign is adopted if  $\theta^* \le \min[\pi, \arccos(-{\beta_e / \beta_\phi})]$, and ``$-$" sign  otherwise.  Notice that  if the boost factor of mediator is very large, a fixed $\theta$ corresponds to two values of $\theta^*$ (for instance, $\theta=0$ means either $\theta^*=0$ or $\theta^*=\pi$), so  in the case of $\beta_\phi > \beta_e  $, positron energy  $E_e$ can not be decided uniquely by $\theta$ without knowing anything about $\theta^* $. 

This multiple-value issue does not happen to photons as $\beta_\gamma =1 \ge \beta_\phi $. In the case of photon, the angles $\theta$ and $\theta^{*}$ are straightforwardly related by
\be
\cos\theta = {\displaystyle \frac{\beta_{\phi} + \cos\theta^{*}}{1 + \beta_{\phi} \cos\theta^{*}}}
\;,\;\;\; {\rm whereas} \;\;
\cos\theta^{*} = {\displaystyle \frac{\cos\theta - \beta_{\phi}}{1 - \beta_{\phi} \cos\theta}} \; ,
\label{eq:angles_1}
\ee
which leads to the well known  ``heading light" effect.
%

\section{Frame transformation for  massless positrons}
\label{app:positron}
\noindent
The kinematics of mediator decays in the Galactic frame ${\cal R}$ are also much clearer when the electron mass $m_{e}$ is set to 0.
In the mediator rest frame ${\cal R}^{*}$, positrons are mono-energetic with energy $E_{e}^{*} \equiv {m_{\phi}}/{2}$. Their momenta is $p_{e}^{*} \equiv E_{e}^{*}$ since they are massless.
Following the definitions in  Appendix.~\ref{sec:appendix_A}, a  boost from the rest frame to the Galactic frame yields
\be
p_{x} = \Gamma_{\! \phi} \! \left( p_{x}^{*} + \beta_{\phi} E_e^{*} \right) =
{\displaystyle \frac{m_{\chi}}{2}} \! \left( \cos\theta^{*} + \beta_{\phi} \right)
\; {\rm and} \;
E_{e} = \Gamma_{\! \phi} \! \left( \beta_{\phi} \, p_{x}^{*} + E_e^{*} \right) =
{\displaystyle \frac{m_{\chi}}{2}} \! \left( 1 + \beta_{\phi} \cos\theta^{*} \right) .
\label{eq:Lorentz_Boost_1}
\ee
The velocity $\beta_{e}^{*} = c$ of the massless positron in the rest frame ${\cal R}^{*}$ is larger than the velocity $\beta_{\phi}$ of this frame with respect to the Galactic frame ${\cal R}$. As a consequence, the angle $\theta$ spans all possible values from 0 to $\pi$. Notice that as the mediator $\phi$ is much lighter than the DM species $\chi$, it is ultra-relativistic with velocity $\beta_{\phi}$ in the Galactic frame ${\cal R}$ close to the speed of light $c$. The positron angular distribution $\Dpos(\cos\theta , E_{e})$ is strongly peaked around $\theta = 0$ in ${\cal R}$ even though its rest frame counterpart $\Dpos^{*}(\cos\theta^{*} , E^\star_{e})$ is isotropic in ${\cal R}^{*}$.

Various methods may be used to derive $\Dpos(\cos\theta, E_{e})$ from $\Dpos^{*}(\cos\theta^{*}, E^\star_{e})$.
\noindent
In ${\cal R}^{*}$, the positron angular distribution is given by
\be
dN_{e} = \Dpos^{*}(\cos\theta^{*}) \, d\Omega^{*} \equiv {\displaystyle \frac{d\Omega^{*}}{4 \pi}} = \frac{1}{2} \, d(- \! \cos\theta^{*}) \; .
\ee
Taking the derivative of the second relation in~(\ref{eq:angles_1}) yields
\be
dN_{e} = \frac{1}{2} \, {\displaystyle \frac{d(- \! \cos\theta^{*})}{d(- \! \cos\theta)}} \; d(- \! \cos\theta) \equiv
{\displaystyle \frac{1 - \beta_{\phi}^{2}}{(1 - \beta_\phi \cos\theta)^{2}}} \times {\displaystyle \frac{d\Omega}{4 \pi}} \; .
\ee
The angle $\theta$ in the Galactic frame corresponds to the energy $E_{e}$ such that
\be
E_{e} =
{\displaystyle \frac{m_{\chi}}{2}} \left( 1 + \beta_{\phi} \cos\theta^{*} \right) =
{\displaystyle \frac{m_{\chi}}{2}} \left\{ {\displaystyle \frac{1 - \beta_{\phi}^{2}}{(1 - \beta_\phi \cos\theta)}} \right\} \equiv
{\displaystyle \frac{{m_{\phi}^{2}}/{2 m_{\chi}}}{(1 - \beta_\phi \cos\theta)}} \; .\label{theta:energy}
\ee
We introduce the energy $E_{e}$ with the positron distribution $\Dpos(\cos\theta , E_{e})$ through the appropriate Dirac function and get
\be
d^{2}N_{e} = \Dpos(\cos\theta , E_{e}) \, d\Omega \, dE_{e} \equiv
\frac{1}{4 \pi} \! \times \! {\displaystyle \frac{{m_{\phi}^{2}}/{m_{\chi}^{2}}}{(1 - \beta_\phi \cos\theta)^{2}}} \times
\delta \! \left\{E_{e} - {\displaystyle \frac{{m_{\phi}^{2}}/{2 m_{\chi}}}{(1 - \beta_\phi \cos\theta)}} \right\} \, d\Omega \, dE_{e} \;, 
\ee
from which we obtain the positron angular distribution as equation~\eqref{distr:twobody} in the main text.

For a continuous spectrum of positron energy $E^\star_e$, like in the case of a three-body decay,  one needs to use the general form: 
\be
 {\mathcal D}_e (\cos \theta, E_e)  = {1\over 2\pi} {d^2N_e\over d\cos\theta dE_e} = {1\over 2\pi} {d^2N_e\over d\cos\theta^* dE^*_{\rm e}}\times {  d\cos\theta^* dE_{\rm e}^* \over d\cos\theta  dE_{\rm e}}.
\ee
By recalling that in the center-of-mass frame the scalar mediator decays isotropically, and applying relativity transformation rules, one can simplify this form to:
\be
 {\mathcal D}_e (\cos \theta, E_e)  = {1\over 4\pi}   \left({  d\cos\theta  dE_{\rm e} \over  d\cos\theta^* dE_{\rm e}^* }\right)^{-1} \times {dN_e\over dE^*_{\rm e}} = {1\over 4\pi \Gamma_\phi(1-\beta_\phi \cos\theta) } \times {dN_e\over dE^*_{\rm e}} \, ,
\ee
where the last factor, describing the energy spectrum of a mediator decay at rest, is uniquely decided by the concrete particle model.

\bibliographystyle{apsrev}
\bibliography{reference}
\end{document}